\documentclass[12pt,preprint2]{aastex}
\newcommand{\kms}{km~s$^{-1}$}

\newcommand{\hi}{H~{\sc i}}
\newcommand{\lya}{Ly$\alpha$}
\newcommand{\fuse}{{\it FUSE}}
\newcommand{\hst}{{\it HST}}
\newcommand{\iue}{{\it IUE}}
\newcommand{\QTR}[2]{\csname text#1\endcsname{#2}}

\begin{document}

\title{A Composite Extreme Ultraviolet QSO Spectrum from \fuse}

\author{Jennifer E.\ Scott\altaffilmark{1}, Gerard A.\ Kriss\altaffilmark{1,2},
Michael Brotherton\altaffilmark{3},
Richard F.\ Green\altaffilmark{4}, John Hutchings\altaffilmark{5},
J.\ Michael Shull\altaffilmark{6}, \& Wei Zheng\altaffilmark{2}}

\altaffiltext{1}{Space Telescope Science Institute, 3700 San Martin Drive,
Baltimore, MD  21218 USA; [jescott,gak]@stsci.edu}

\altaffiltext{2}{Center for Astrophysical Sciences, Department of Physics and Astronomy,
The Johns Hopkins University, Baltimore, MD 21218 USA; zheng@pha.jhu.edu}

\altaffiltext{3}{University of Wyoming, Department of Physics and Astronomy,
Laramie, WY, 82071 USA;mbrother@uwyo.edu}

\altaffiltext{4}{Kitt Peak National Observatory, National Optical Astronomy Observatories,
P.O. Box 26732, 950 North Cherry Avenue, Tucson, AZ 85726  USA;
green@noao.edu}

\altaffiltext{5}{Herzberg Institute of Astrophysics, National
Research Council Canada, Victoria, BC V9E 2E7, Canada; john.hutchings@hia.nrc.ca}

\altaffiltext{6}{Center for Astrophysics and Space Astronomy,
Department of Astrophysical and Planetary Sciences, University of Colorado,
Boulder, CO 80309 USA;
mshull@casa.colorado.edu}

\addtocounter{footnote}{-6}

\begin{abstract}

The {\it Far Ultraviolet Spectroscopic Explorer} (\fuse) has surveyed a large
sample ($ > 100$) of active galactic nuclei in the low-redshift
universe ($z < 1$).  Its response at short wavelengths
makes it possible to measure directly
the far ultraviolet spectral properties of quasistellar objects (QSOs) 
and Seyfert 1 galaxies at $z < 0.3$.
Using archival
\fuse\ spectra, we form a composite extreme
ultraviolet (EUV) spectrum of QSOs at $z \le 0.67$.
After consideration of many possible sources of
systematic error in our analysis, we find that
the spectral slope of the \fuse\ composite spectrum,
$\alpha= -0.56^{+0.38}_{-0.28}$ for $F_{\nu} \propto \nu^{\alpha}$, is significantly
harder than the EUV ($\lambda \lesssim 1200$ \AA) portion of the composite spectrum of
QSOs with $z>0.33$ formed from archival {\it Hubble Space Telescope} spectra,
$\alpha=-1.76 \pm 0.12$.
We identify several prominent emission lines in the \fuse\ composite and
find that the high-ionization O~{\sc vi} and Ne~{\sc viii} emission lines
are enhanced relative to the \hst\ composite.
Power law continuum fits to the individual \fuse\ AGN spectra
reveal a correlation between EUV spectral slope and AGN luminosity 
in the \fuse\ and \fuse+\hst\ samples in the sense that lower luminosity
AGNs show harder spectral slopes. 
We find an anticorrelation 
between the hardness of the EUV spectral slope and
AGN black hole mass, using estimates of this quantity found in the literature.
We interpret these results in the context of the well-known anticorrelation
between AGN luminosity and emission line strength, the Baldwin effect,
given that the median luminosity of the \fuse\ AGN sample is
an order of magnitude lower than that of the \hst\ sample.
 
\end{abstract}

\section{Introduction}

The ubiquity with which QSOs display spectral properties such as
power law continua and broad emission lines
over wide ranges in
luminosity and redshift has led to the use of composite spectra
to study their global properties.
In this paper, we present a composite far, or extreme, ultraviolet (EUV)
spectrum of low-redshift AGNs.
Information about the continuum in the rest-frame
ultraviolet is particularly critical
for understanding the formation of the emission lines, for 
characterizing the UV bump in QSO spectral energy distributions, 
and for determining the ionization state of the intergalactic medium (IGM).
Composite QSO spectra covering the rest-frame ultraviolet
have been constructed for AGNs with
$0.33 < z < 3.6$ from \hst\ (Zheng et al.\ 1997; 
Telfer et al.\ 2002, Z97 and T02 hereafter),
and at $z > 2$ from ground-based samples like the
Large Bright Quasar Survey (Francis et al.\ 1991), the
Sloan Digital Sky Survey (SDSS, Vanden Berk et al.\ 2001), and
the First Bright Quasar Survey (Brotherton et al.\ 2001).
Z97 and T02 reported that
the \hst\ composite spectrum, which covers rest wavelengths $\sim$350-3000 \AA, 
shows a spectral break in the power law continuum at 1050-1300 \AA\
in the sense that the EUV spectral shape blueward of the break 
is softer than the slope redward of the break, in the  
near ultraviolet (NUV).  In this composite, they identify
several prominent far-UV emission features including lines 
due to Ly$\alpha$ blended with the
N~{\sc v} doublet, Ly$\beta$ blended with the O~{\sc vi} doublet,
and Ne~{\sc viii}.

The bandpass of the {\it Far Ultraviolet Spectroscopic Explorer} (\fuse) 
(Moos et al.\ 2000; Sahnow et al.\ 2000),
905-1187 \AA, allows us to
examine the EUV properties of local AGNs.
Therefore, we can study the same rest-frame EUV wavelength region covered by
the \hst\ composite spectra, for AGNs with redshifts less than $0.33$.
The low redshifts of these \fuse\ AGNs ensure that although
the \fuse\ aperture limits it to observing relatively bright AGNs,
our sample contains a substantial fraction of intrinsically low-luminosity
($\log(\lambda L_{\lambda} [\mathrm{erg\: s^{-1}}] )$ at 1100~\AA\ $\equiv \log(\lambda L_{1100}) < 45$) AGNs.
The \fuse\ sample combined with the \hst\ sample yields a sample of AGNs
with a spread of nearly five orders of magnitude in luminosity from which
we can investigate trends in EUV spectral shape with luminosity invoked to
account for the anticorrelation between AGN luminosity and 
emission line equivalent width, the Baldwin effect
(Baldwin 1977; Netzer, Laor, \& Gondhalekar 1992; Green 1996, 1998;
Wang, Lou, \& Zhou 1998; Dietrich et al.\ 2002),
and the dependence of the strength of the
Baldwin effect on the ionization potential of the emitting ion
(Zheng, Fang, \& Binette 1992; Zheng \& Malkan 1993; 
Zheng, Kriss, \& Davidsen 1995;
Espey \& Andreadis 1999;
Dietrich et al.\ 2002; Kuraszkiewicz et al.\ 2002; Shang et al.\ 2003).

The AGNs in our \fuse\ archival sample all have redshifts less than $0.7$, affording
the advantage that the determination of the mean EUV spectral
index requires a less significant correction for IGM absorption than was
required for the \hst\ sample.
After consideration of various possible systematic effects on the
analysis,
we compare the spectral shape of the composite and the strength of the emission lines
to those found in the EUV spectrum of AGNs with
$z > 0.33$ compiled from \hst\ data by T02.  We also fit power law
continua to each individual AGN spectrum in the \fuse\ sample and
examine the results for correlations of the spectral slope with redshift and luminosity.
Finally, we compile estimates of black hole mass for several of the sample AGNs
to test for an anticorrelation between this quantity and spectral slope.
Such a correlation is
a prediction of an evolutionary model in which the central black holes
of AGNs accrete mass over time
and the peak of the UV bump in the spectral energy distribution shifts
to longer wavelengths, resulting in softer ionizing continua in the UV and
soft X-rays for higher mass/luminosity AGNs (Wandel 1999a,b).  In such a scenario
we would also expect to find a correlation between UV spectral slope and
the accretion disk temperature, as estimated from 
the AGN luminosity and black hole mass, and we investigate this with
the \fuse\ data as well.

\section{The Sample}

From the \fuse\ archives, we downloaded the 165 spectra of AGNs with $z < 1$ 
that were public as of 2002 November.
We processed the raw data using standard 
\fuse\ calibration pipelines (see
Sahnow et al.\ 2000) to extract the spectra, to perform background subtraction using
updated background models and subtraction algorithms, and
to perform wavelength and flux calibrations.  We also implement a correction
for the $``$worm", a dark stripe running in the dispersion direction
on one of the \fuse\ detector segments.
See the \fuse\ web pages\footnote{http://fuse.pha.jhu.edu} 
for details about this feature.

Following a procedure similar to that of T02, we 
excluded spectra of broad-absorption-line quasars and spectra with 
signal-to-noise ratios (S/N) $\lesssim 1$
over large portions.
We also exclude spectra of AGNs with strong narrow emission lines, e.g.\ NGC 4151,
strong stellar features in their continua, e.g.\ NGC 7496, or 
strong interstellar molecular hydrogen absorption, e.g.\ 1H 2107-097.
A total of 128 spectra of 85 AGNs, all with $z \le 0.67$,
meet the criteria for inclusion in the sample.
The 85 AGNs in the \fuse\ sample are listed in Table~\ref{table-alpha}, along
with their redshifts and Galactic reddening values from Schlegel, Finkbeiner, \& Davis (1998).

\section{Composite Spectrum Construction}
We follow the same procedure as T02 for the construction of the composite spectrum. 
The reader should consult that paper for a detailed
discussion of the corrections applied to each sample spectrum, which we summarize here.  
\begin{enumerate}
\item We correct the spectrum for Galactic extinction using the 
Cardelli, Clayton, \& Mathis (1989) extinction curve,
$E(B-V)$ values listed in 
Table~\ref{table-alpha} (Schlegel et al.\ 1998), 
and $R_{V}=3.1$.

\item We exclude wavelength regions affected by interstellar absorption lines.

\item We correct for Lyman limit absorption using $\tau_{\rm LLS}=\ln{(F_{+}/F_{-})}$,
if the S/N
below the Lyman break is greater than one,
where $F_{+}$ is the median flux in selected windows redward of the Lyman break and
$F_{-}$ is the median flux blueward of the break.

\item We apply a statistical correction for the line of sight absorption due
to the \lya\ forest and the Lyman valley (M\o ller \& Jakobsen 1990).

\item We shift the AGN spectrum to the rest frame.

\item We resample the spectrum to common 1 \AA\ bins.

\end{enumerate}

The lower redshifts of our sample AGNs compared with
the \hst\ sample of T02 compels us to
use different parameters to correct for
\lya\ forest absorption mentioned above,
so we discuss them briefly here.  Like T02, we use the distribution of absorbers
given by 

\begin{equation}
\frac{\partial^{2}n}{\partial z \partial N} \propto (1+z)^{\gamma} N^{-\beta}.
\label{equ:dndzdN}
\end{equation}

We account for column densities in the range $12.2 < \log N < 16.7$.
For the column density distribution parameter, we use the result
found by Dav\'{e} \& Tripp (2001) from {\it HST}/Space Telescope Imaging Spectrograph (STIS) 
echelle spectra of 
two QSOs at $z\sim0.3$, 
$\beta=2.0$ for $12.2 < \log{N} < 14.4$. 
For $14.4 < \log{N} < 16.7$, we use $\beta=1.35$ from the {\it HST}/Goddard High Resolution
Spectrograph study by
Penton et al.\ (2000).  
For the redshift distribution parameter, we use $\gamma=0.15$  (Weymann et al.\ 1998).
We normalize the distribution of absorbers in Equation~\ref{equ:dndzdN} 
by $1.34 \times 10^{-11}$ cm$^{2}$ at $\log{N}=13$ and $z=0.17$ and assume
a Doppler parameter of 21 \kms\ (Dav\'{e} \& Tripp 2001).

\begin{figure}
\epsscale{0.55}
\plotone{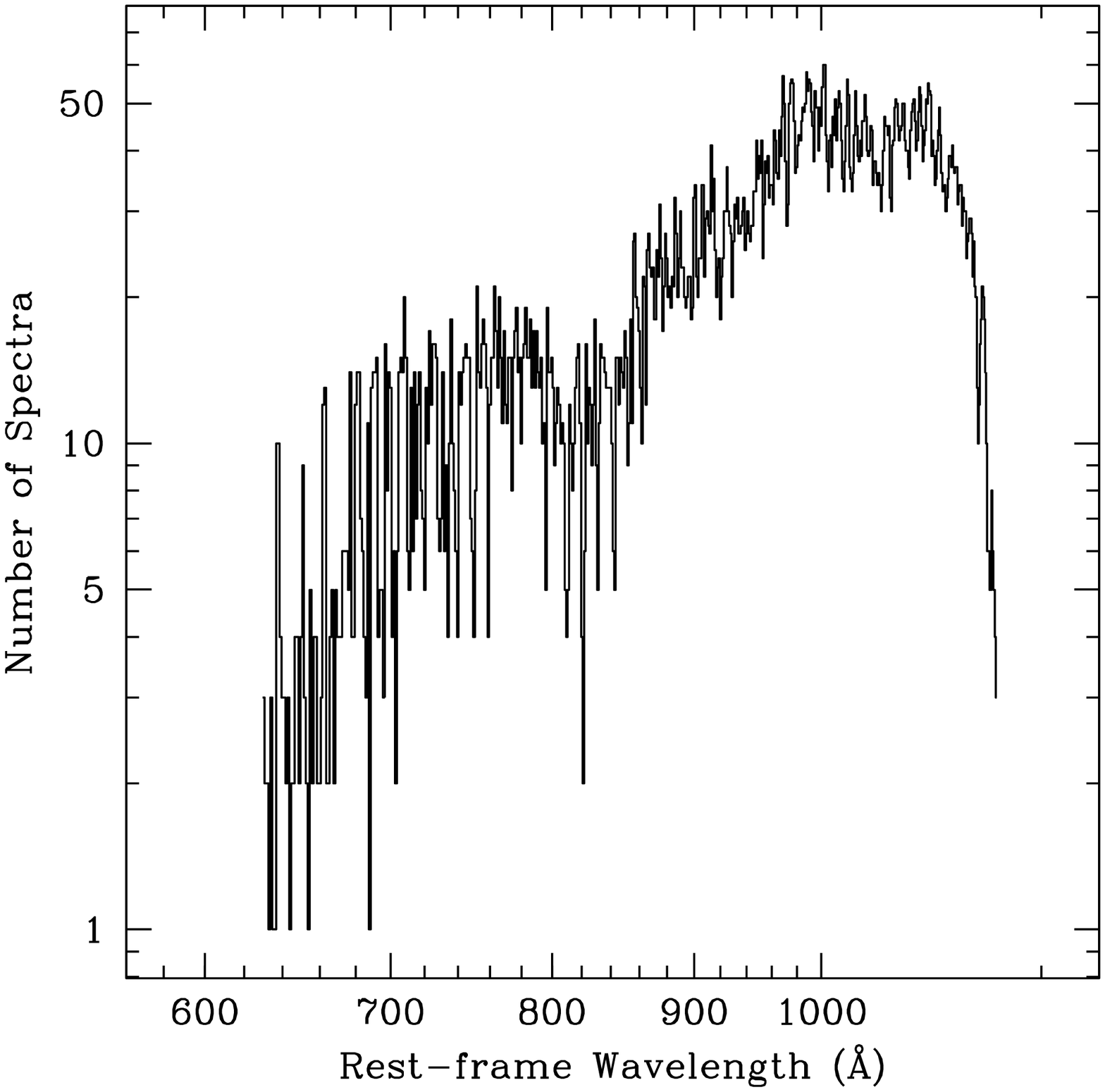}
\figcaption{Number of spectra used in the composite spectrum as a function of rest wavelength.
\label{fig:nqso}}
\plotone{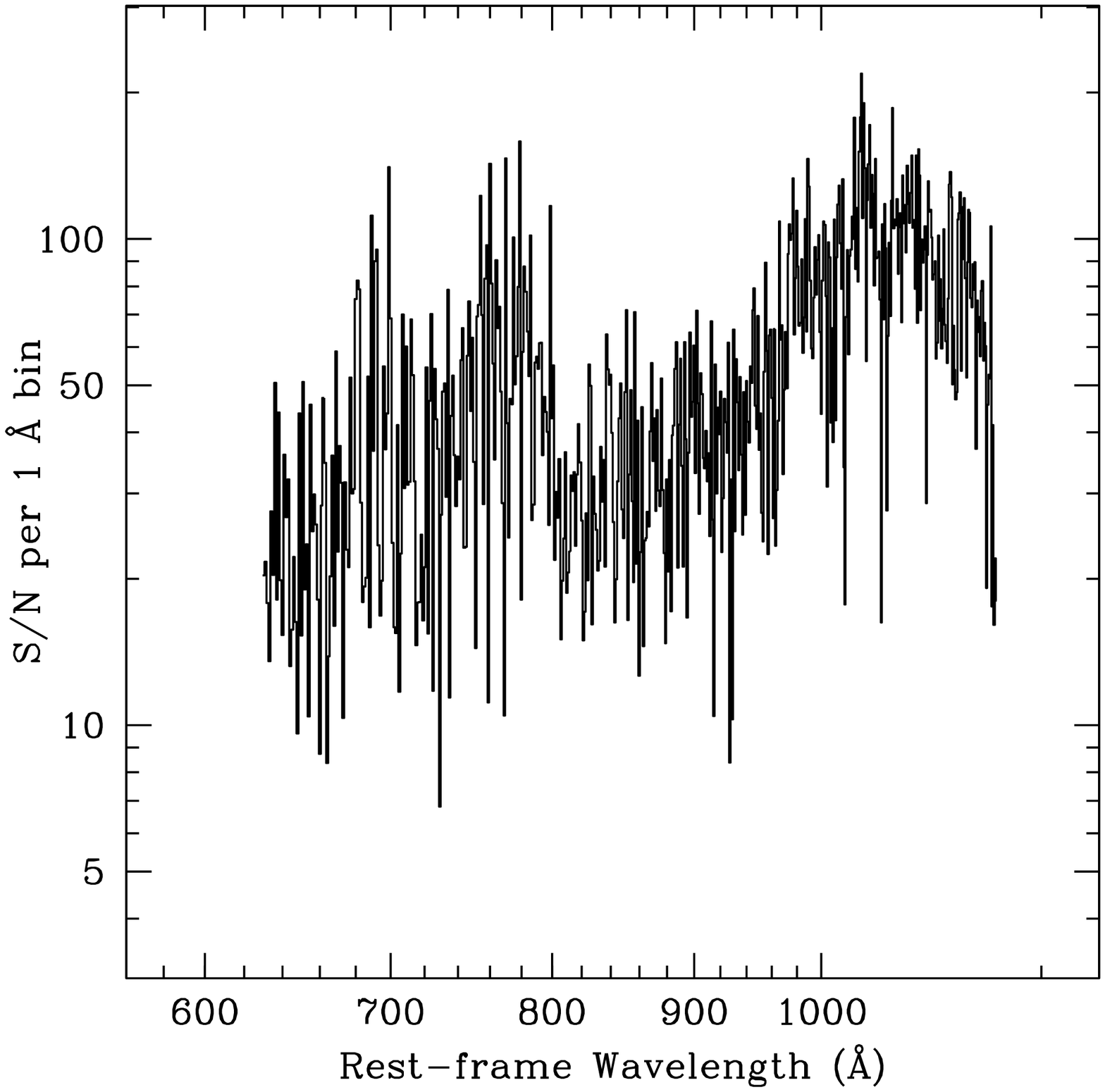}
\figcaption{Signal-to-noise ratio in the composite spectrum per 1 \AA\ bin.
\label{fig:sn}}
\epsscale{1.0}
\end{figure}

We combine the sample spectra using the bootstrap technique described by T02.
To summarize briefly, we begin the bootstrap procedure at the central portion of
the output composite, specifically the region between 850 and 950 \AA.
We then include spectra that fall at longer wavelengths  in
sorted order to longer wavelengths. Finally, we include those at shorter wavelengths
in sorted order to shorter wavelengths.  The overall composite is renormalized at 
each step.
Figures~\ref{fig:nqso} and \ref{fig:sn} show the number of spectra contributing
to the final composite in each wavelength bin, and the S/N per wavelength bin
in the final composite, respectively.  The non-smooth appearance of these two histograms
is caused by the large number of discrete wavelength regions surrounding interstellar 
absorption lines we omitted from the calculation. 

\begin{figure}
\plotone{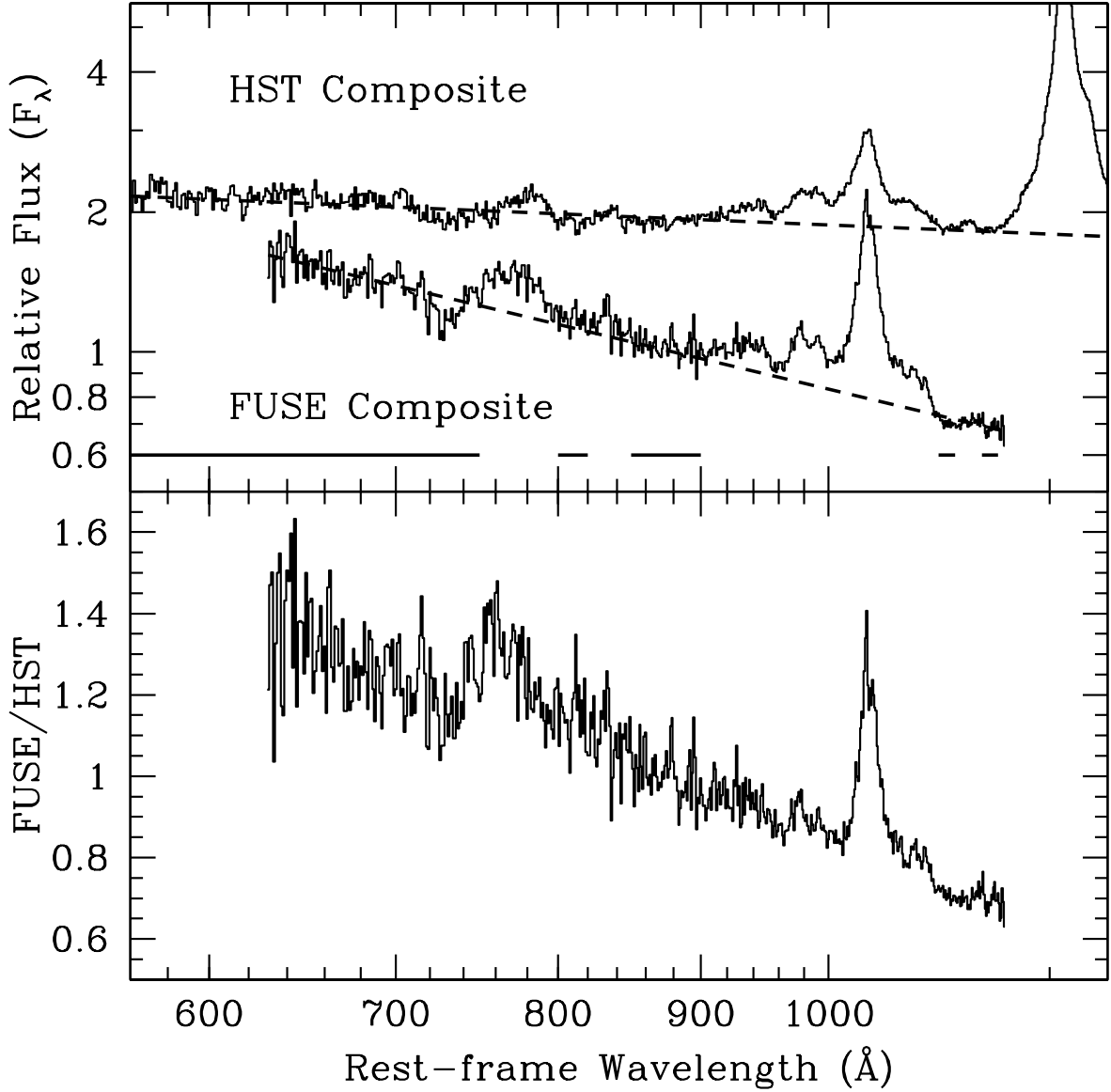}
\figcaption{{\it Top panel:} Composite AGN spectrum with power law continuum fit 
({\it dashed lines}) and
wavelength regions used in fit ({\it solid line segments});
\hst\ composite from Telfer et al.\ (2002) shown for comparison.
{\it Bottom panel:}  Ratio of \fuse\ to \hst\ composite spectra.
\label{fig:comp}}
\end{figure}

The final \fuse\ composite spectrum, shown in the top panel of Figure~\ref{fig:comp},
covers the rest wavelength range 630-1155 \AA.
We used the IRAF task {\it specfit} (Kriss 1994) to fit a power law of
the form  $F_{\nu} \propto \nu^{\alpha}$
to the continuum of the composite using wavelength regions
free of emission lines:
630-750, 800-820, 850-900, 1095-1100, and 1135-1150 \AA\ (T02).
The best-fit
power law index is $\alpha= -0.56$, and we show this fit with the dashed line in Figure~\ref{fig:comp}. 
We compare this index to the values of $\alpha_{\rm EUV}$ from continuum fits to the
\hst\ composite for wavelengths $>500$ \AA\ in T02.
In the top panel of Figure~\ref{fig:comp},
we also show the \hst\ composite from
T02, and in the bottom panel of the figure, we show
the ratio of the \fuse\ and \hst\ composites.  The \fuse\ composite
is significantly harder and shows enhancement of several emission lines,
which will be discussed further in Section~\ref{sec-emlines}. 

\section{Uncertainties}
\label{sec-uncertain}

Because the S/N in the \fuse\ composite is high (Figure~\ref{fig:sn}), the 
statistical error in the continuum fit is small.  The 
largest sources of error are likely to come from cosmic variance
in the spectral shapes of individual AGNs and from systematic
errors in the method we have used to correct the sample spectra
for Galactic and intergalactic absorption.
We estimate the error arising from the range of
spectral shapes of the individual
AGNs that constitute our \fuse\ sample by
creating 1000 bootstrap samples with replacement from the original sample.
From the bootstrap samples, we 
find a standard deviation in $\alpha$ of 0.11.

We now explore how a number of possible systematic errors within our analysis would 
affect the results.
The bootstrap combination technique is robust, so varying it
has little effect on the resulting composite.
If we perform the bootstrap from long wavelengths to
short, $\alpha$ changes by $+0.03$.
Normalizing the spectrum at different wavelength
intervals, 1050-1150 \AA\ (T02) or 800-900 \AA\ versus the fiducial range 850-950 \AA\
changes $\alpha$ by an amount less than the statistical error in the power
law fit from {\it specfit},
$+0.01/-0.003$.

We find that the results are sensitive to the extinction
correction in terms of both the individual values of $E(B-V)$ and the
adopted ratio of total-to-selective extinction, $R_{V}$. 
Changing all individual values of $E(B-V)$ by $\pm 1\sigma$, where we 
estimate $1\sigma=0.16 E(B-V)$ (Schlegel et al.\ 1998) changes 
$\alpha$ by $\pm 0.16$.  For our fiducial correction for Galactic
reddening, we used 
the average value of $R_{V}$ for sightlines
through the diffuse interstellar medium of the Milky Way, 3.1 
(Cardelli, Clayton, \& Mathis 1989).
However, for individual lines of sight, it may vary as widely as $R_{V}=2.5$ to
$R_{V}=5.5$ (Clayton \& Cardelli 1988).  If instead of $R_{V}=3.1$, we use
instead $R_{V}=2.8$ or $R_{V}=4.0$ in the extinction law,  the spectral
index of the composite spectrum changes by $+0.06$ and $-0.19$, respectively.

The composite is also sensitive to the value of the
column density distribution parameter, $\beta$. 
Because the result of Dav\'{e} \& Tripp (2001) is larger than
typically inferred for $\beta$ from lower resolution data
(Penton et al.\ 2000; Dobrzycki et al.\ 2002) 
or at high redshift (Kim, Cristiani, \& D'Odorico 2001),
we reduce it from the 
fiducial value of 2.0 to 1.5.  This increases
$\alpha$ by 0.3.  
The composite is fairly insensitive to the redshift
distribution parameter, $\gamma$, which we varied from 0.15 
(Weymann et al.\ 1998) to 0.54 (Dobrzycki et al.\ 2002).
Raising and lowering the upper column density limits for
the \lya\ forest correction to $\log N =16.9$ and $\log N = 16.5$ 
changes $\alpha$  by $+0.06$ and $-0.04$, respectively.

\begin{figure}
\plotone{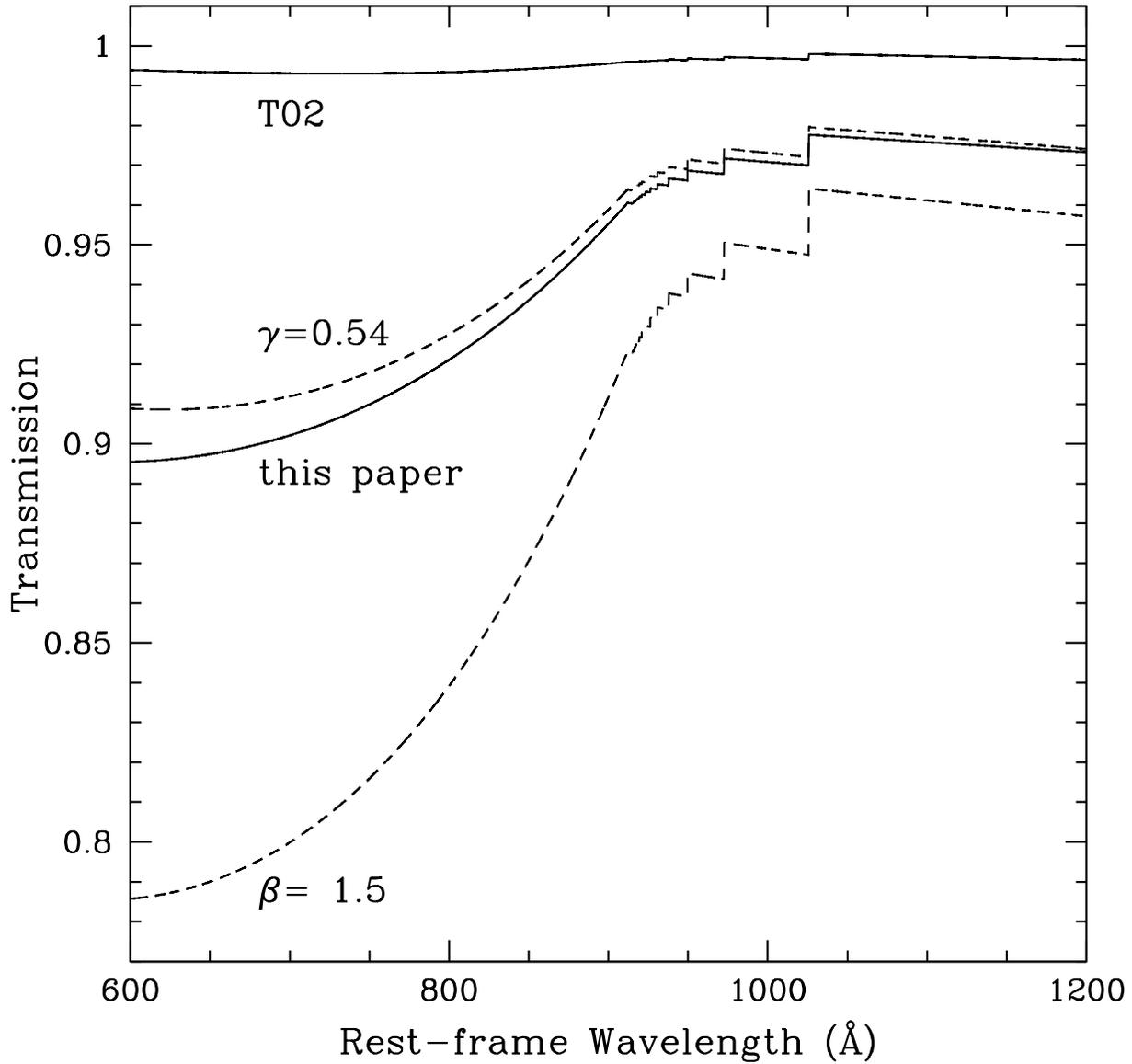}
\figcaption{
Lyman valley correction for a $z=0.1$ AGN with the
parameters used by T02, and the correction used here, where $\gamma=0.15$, $\beta=2.0$
in Equ.~\ref{equ:dndzdN}.
The dashed lines show how the correction changes for different $\gamma$ and $\beta$. 
\label{fig:valley}}
\end{figure}

The relatively strong effect that varying $\beta$ has on the results is somewhat
surprising, since at the low redshifts considered here,
the number of \lya\ forest absorbers in any given spectrum is small.
At the far ultraviolet wavelengths relevant to this composite
spectrum, we have corrected also for line blanketing in the Lyman series
and for Lyman continuum absorption, the Lyman valley (M\o ller \& Jakobsen 1990), 
using the same \lya\ forest parameters as
discussed above. The effect that changing these parameters has on the Lyman valley
correction for a $z=0.1$ AGN 
is shown in Figure~\ref{fig:valley}.  We also show here the correction
calculated using the \lya\ forest parameters of T02.
This correction is less than 1\% over the whole spectrum because, relative
to the shallow power law number distribution observed at low redshift, T02
used $\gamma=2.46$ appropriate for the high redshifts of their \hst\ AGN sample.
Extrapolating that power law to $z=0.1$ with the same normalization results in
a factor of $\sim$50 underestimate in the number of \lya\ absorbers at $z = 0.1$.  
In this figure,
we can see the effect that changing $\gamma$ and $\beta$ has on the correction
used for the \fuse\ data.  Changing $\gamma$ from 0.15 to 0.54 has relatively little effect,
while increasing $\beta$ from 2.0 to 1.5 
increases the correction substantially, to $\gtrsim$20\%.
It should be noted, however, that 
in any given AGN spectrum, no identifiable Lyman valley trough is observed.  Deriving a
composite spectrum with no correction for the \lya\ forest and
Lyman valley results in $\alpha=-0.75$, a marginally softer index than quoted
above, though
still significantly harder than the EUV spectral index derived from
the \hst\ data by T02.

Adding all the systematics discussed in this section in quadrature,
we estimate the total uncertainty in $\alpha$ to be $(+0.38,-0.28)$.
The spectral index from the \hst\ composite, $\alpha=-1.76 \pm 0.12$, 
differs from the \fuse\ value, 
$\alpha=-0.56$, by
$\sim$3 times this amount.

\section{Redshift, Luminosity, and Reddening Subsamples}
\label{sec-lz}

Here we examine trends in the \fuse\ AGN sample with redshift,
luminosity, and Galactic reddening using composites formed 
from various subsamples of the full AGN sample.
The subsamples discussed in this section are
summarized in Table~\ref{table-subsamples}.

\begin{figure}
\epsscale{0.55}
\plotone{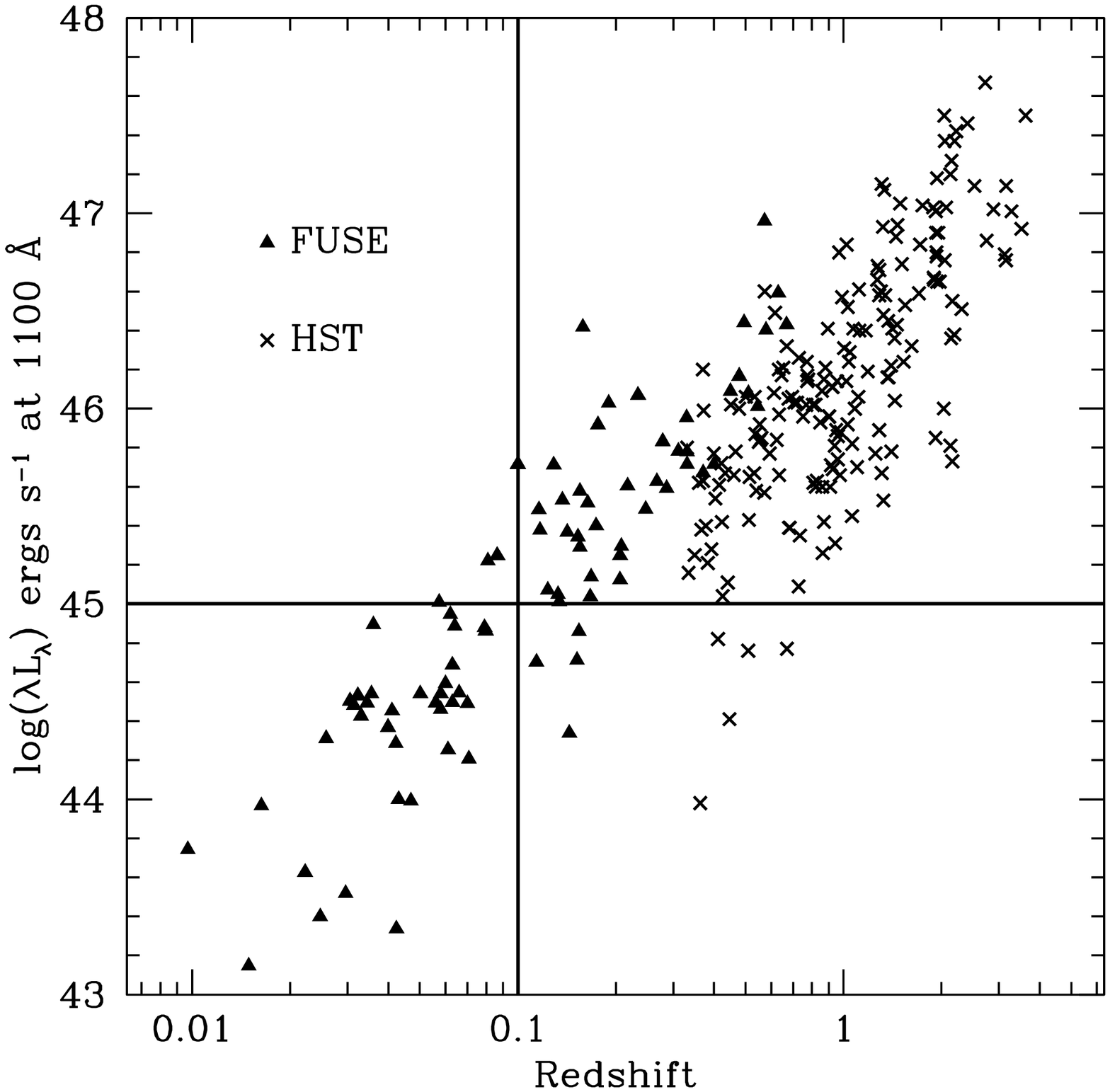}
\figcaption{Luminosity versus redshift for \fuse\ and \hst\ AGN
with lines marking median redshift ($z=0.10$) and median luminosity ($\log \lambda L_{1100}=45.0$).
\label{fig:lz}}
\plotone{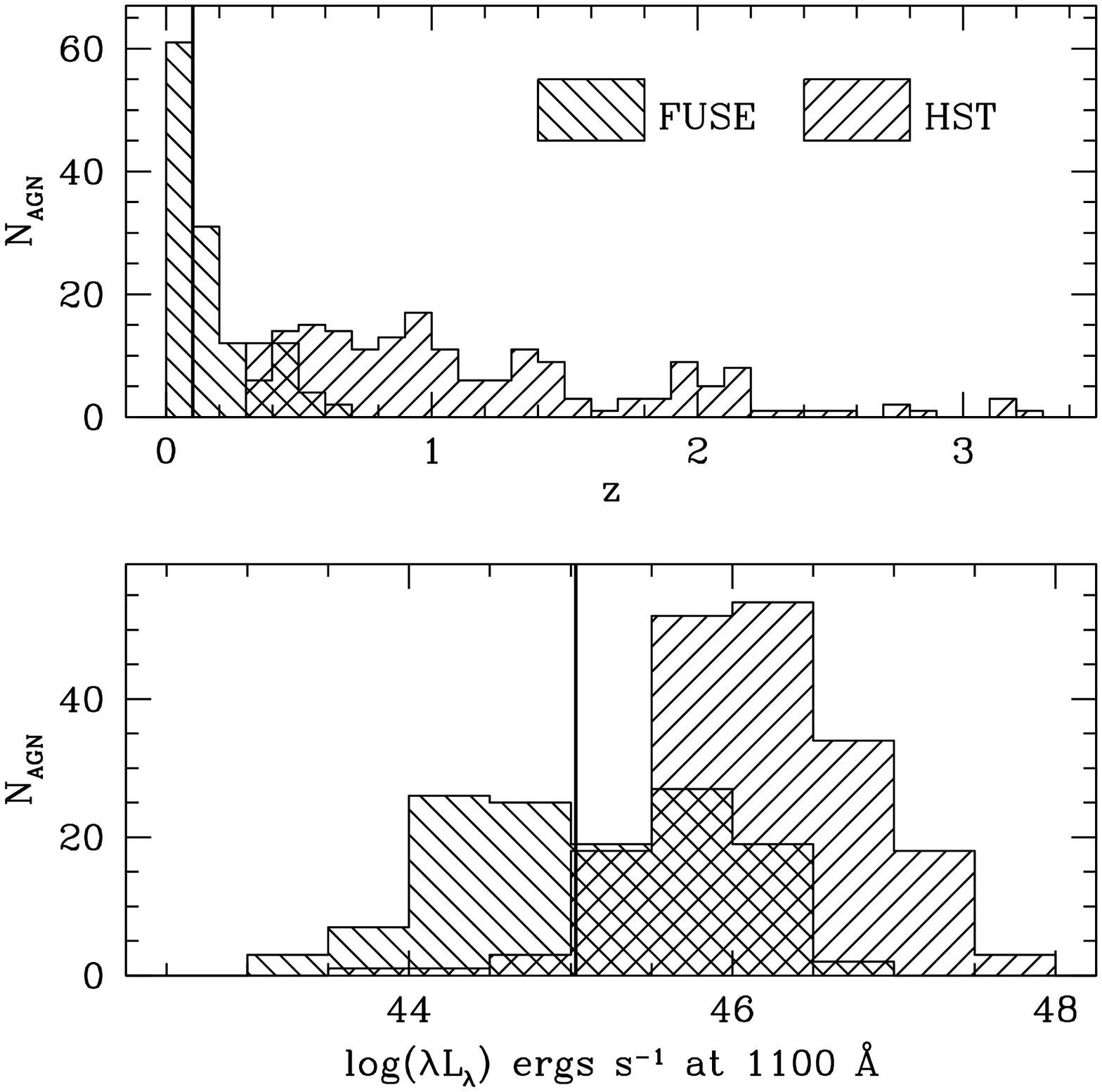}
\figcaption{Histograms of redshifts and luminosities of \fuse\ and \hst\ AGN
with vertical lines marking median \fuse\ redshift and luminosity.
\label{fig:hist}}
\epsscale{1.0}
\end{figure}

First, we calculate the luminosity of each \fuse\ AGN
using the prescription described by T02:
\begin{equation}
L_{\lambda} (1100~\mathrm{\AA}) = F_{c} (1100~\mathrm{\AA})  
\frac{\sum_i F_{s,i}}{\sum_i F_{c,i}} 4 \pi D_L^2,
\label{equ:lum}
\end{equation}
where $F_{c}$ is the flux in the composite, $F_{s}$ is the  flux in the individual spectrum, 
and the sum is performed over the spectral regions where the individual AGN spectrum
overlaps with the composite.
The term $D_L$ is the luminosity distance to the AGN.
These luminosities are listed in Table~\ref{table-alpha}.
We assume 
$\Omega_{0}=1$  and $H_{0}=60$ km~s$^{-1}$ Mpc$^{-1}$, as T02 did.
We show the distribution in redshift and luminosity
of the AGNs in the \fuse\ sample and in the \hst\ sample of T02
in Figure~\ref{fig:lz} with
the median redshift and luminosity of the \fuse\ sample, $z=0.10$ and 
$\log(\lambda L_{1100})=45.0$, 
marked by the vertical and horizontal lines.
Figure~\ref{fig:hist} shows these distributions in 
histograms. 
The redshift histogram illustrates that the \fuse\ sample is
highly concentrated around the median redshift.  The luminosity
histogram shows the large
span of UV continuum luminosity in the combined \fuse\ + \hst\ sample.

\begin{figure}
\epsscale{0.55}
\plotone{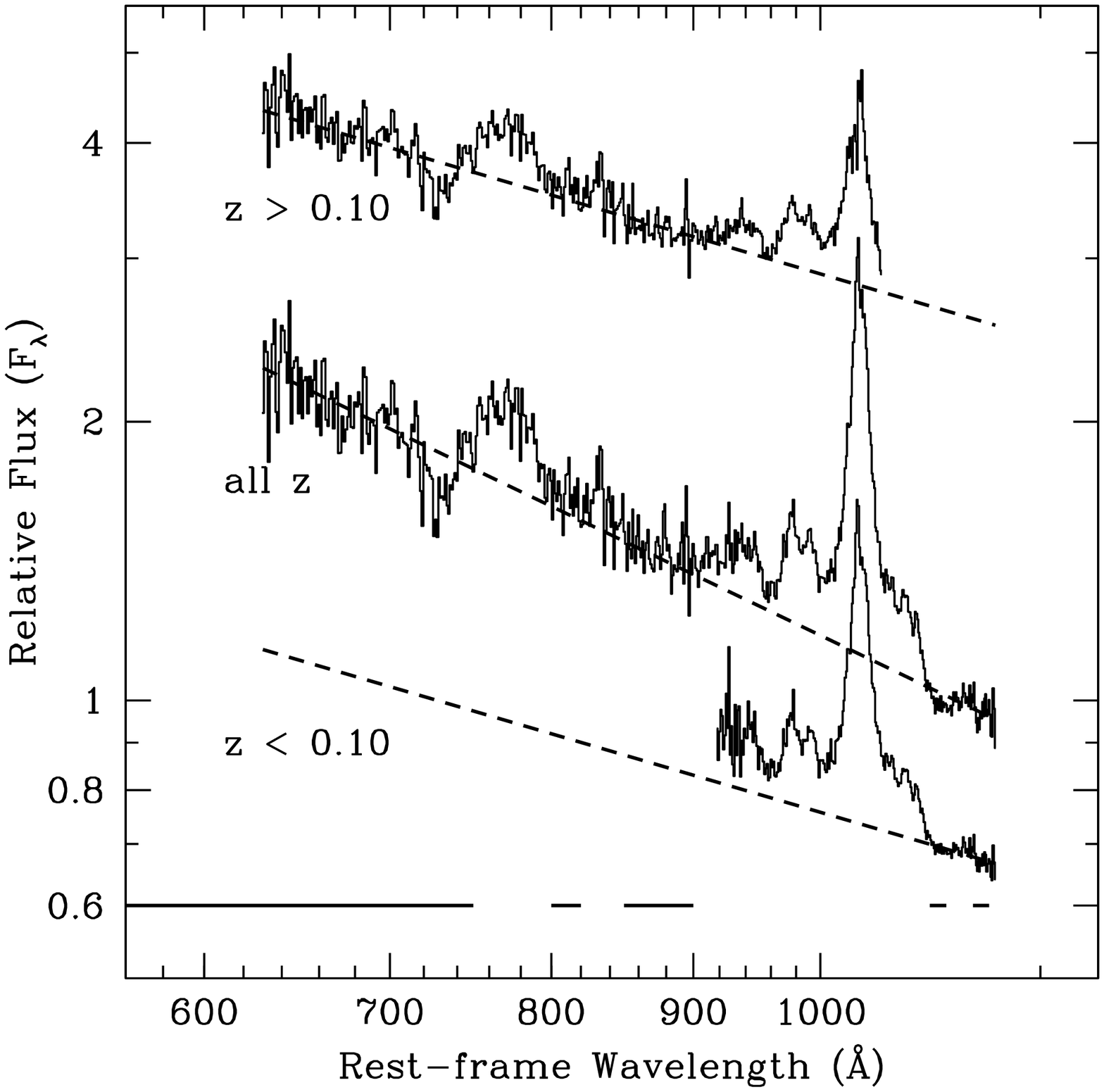}
\figcaption{Composites made from spectra of AGNs with $z < 0.10$ and $z > 0.10$ with
power law continuum fits ({\it dashed lines}) and
wavelength regions used in fit ({\it solid line segments}).
Overall composite with power law continuum
fit shown for comparison.  
\label{fig:compz}}
\plotone{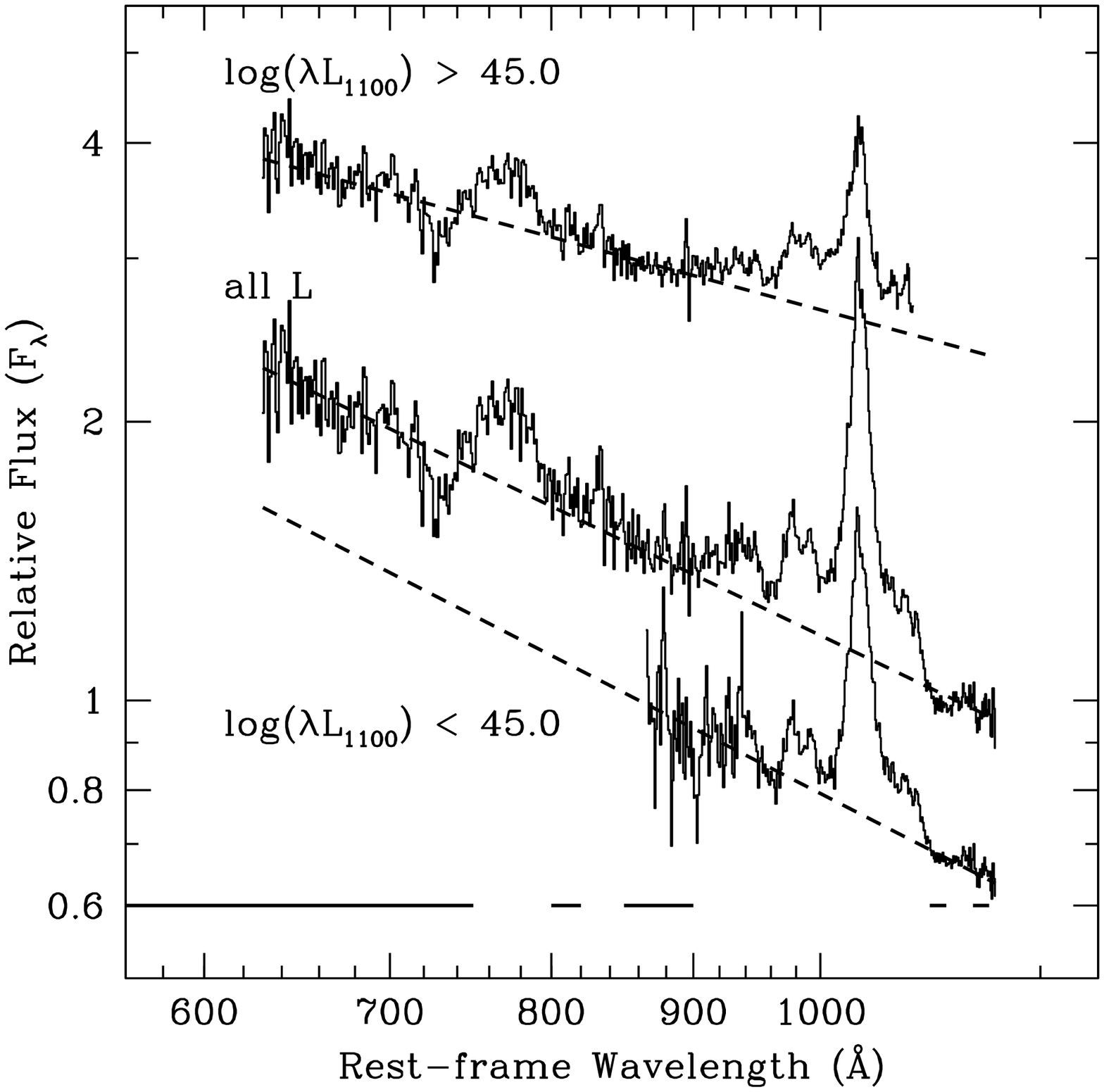}
\figcaption{Composites made from spectra of AGNs with $\log(\lambda L_{1100}) < 
45.0$ and $\log(\lambda L_{1100})> 45.0$ with
power law continuum fits ({\it dashed lines}) and
wavelength regions used in fit ({\it solid line segments}).
Overall composite with power law continuum
fit shown for comparison.
\label{fig:compl}}
\epsscale{1.0}
\end{figure}

Dividing the sample in redshift at $z=0.10$ and constructing
composite spectra from the two redshift subsamples,
we find the same value of $\alpha$ for both low-redshift and high-redshift
composites, $\alpha=-0.74$.  With this redshift cut,
the total composite spectrum is dominated by low-redshift AGNs at
$\lambda \gtrsim 950$ \AA\ and by high redshift AGNs at $\lambda \lesssim 950$ \AA.
Figure~\ref{fig:compz} shows the composites made from all AGNs, and from high-z
and low-z AGNs.

Dividing the sample in luminosity at $\log(\lambda L_{1100})=45.0$
gives $\alpha=0.19$ for low-luminosity 
AGNs and $\alpha=-0.84$ for high-luminosity AGNs. 
Figure~\ref{fig:compl} shows the composites made from all AGNs, and from high-luminosity
and low-luminosity AGNs.
Similar to the low- and high- redshift subsamples, 
the high- and low- luminosity subsamples cover different spectral
ranges.  Only a small spectral region is common to both subsamples: 
866-901~\AA.
The difference between the continuum slopes of the low-luminosity and
high-luminosity composites may come from the limited limited spectral 
range used in the fit in each case.
However, the difference between these two slopes is $2.7\sigma^{+}$,
where $\sigma^{+}$ is the overall uncertainty in the spectral index of the 
\fuse\ composite discussed in Section~\ref{sec-uncertain}.
The low-luminosity composite has a harder spectral shape, and Figure~\ref{fig:compl}
illustrates that it also shows enhanced O~{\sc vi}/Ly$\beta$ emission. 

\begin{figure}
\epsscale{0.55}
\plotone{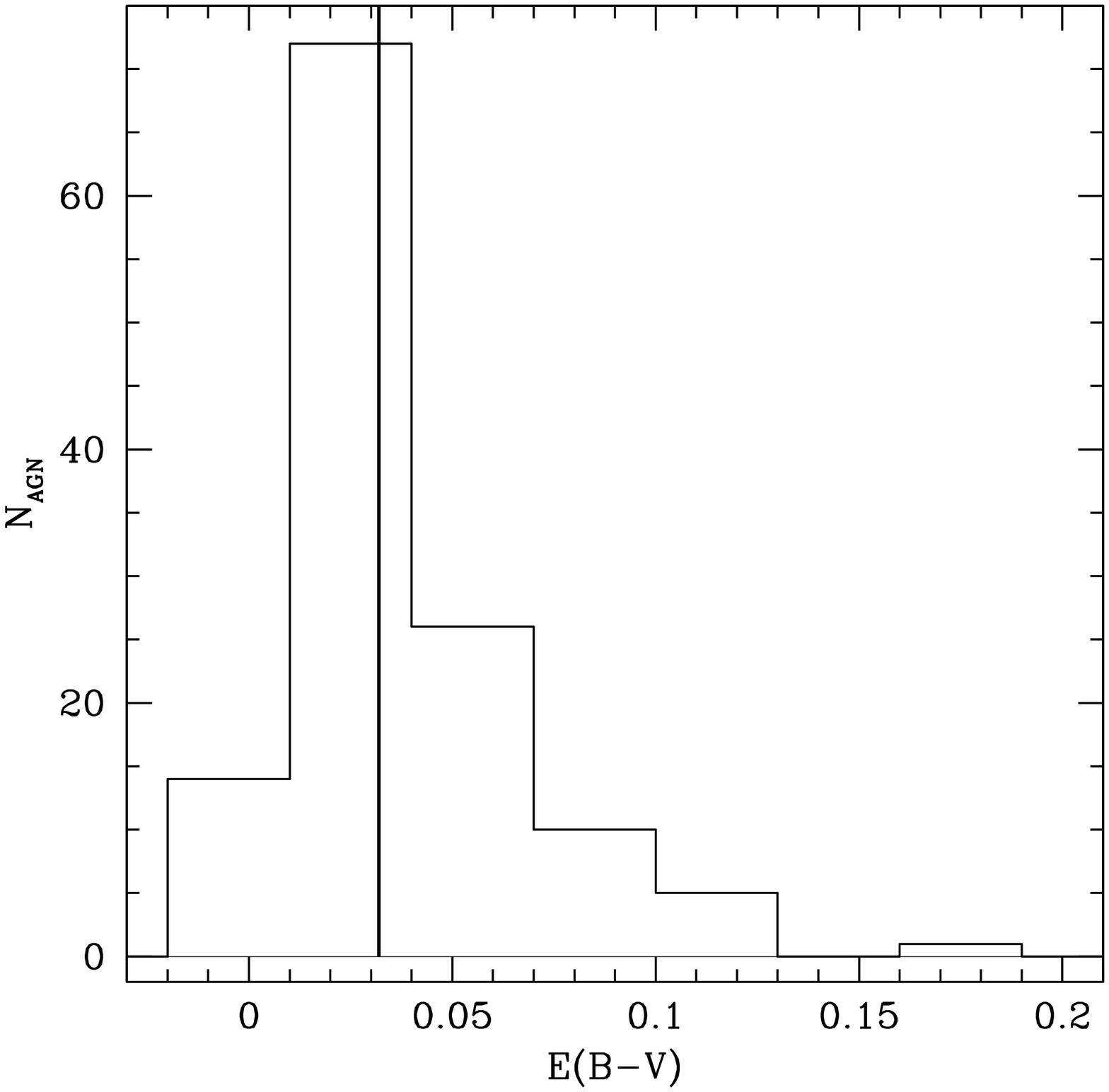}
\figcaption{
Histogram of $E(B-V)$ values for \fuse\ AGNs. The solid line marks the median
$E(B-V)$, 0.032.
\label{fig:ebvhist}}
\plotone{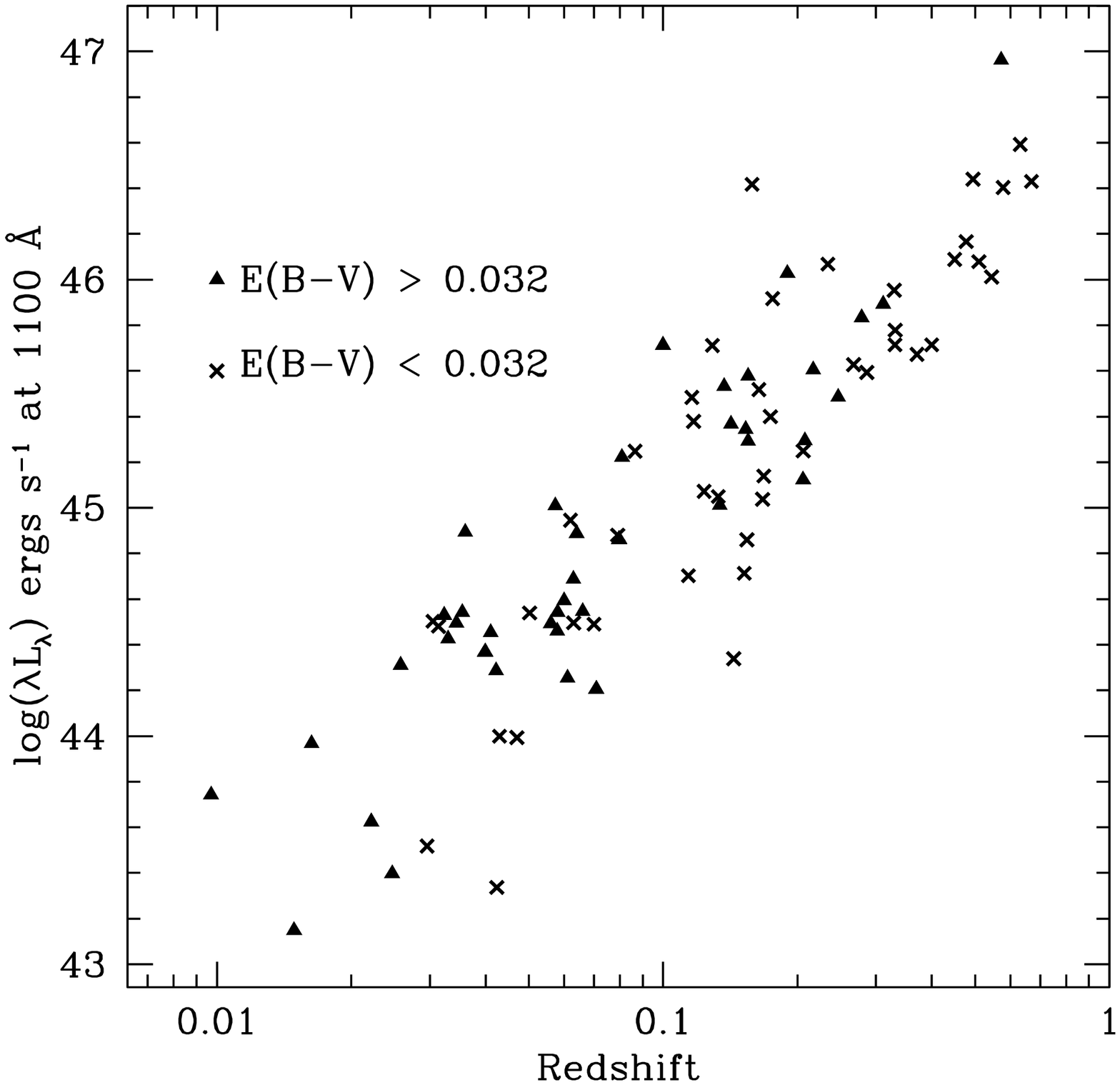}
\figcaption{Distribution in redshift and 1100-\AA\ luminosity for AGNs
with $E(B-V) > 0.032$ ({\it filled triangles}) and AGNs with
$E(B-V) < 0.032$ ({\it crosses})
\label{fig:lzebv}}
\epsscale{1.0}
\end{figure}

The AGNs in the \fuse\ sample lie along sightlines with Galactic reddening
values in the range $0.006 < E(B-V)  < 0.182$.
The median value is $E(B-V)=0.032$.
We show the distribution of reddening values in Figure~\ref{fig:ebvhist}.
Because we found the overall \fuse\ composite to be sensitive to the
reddening correction, as discussed in Section~\ref{sec-uncertain},
we consider whether a bias is introduced into the composite
spectrum by over- or under-correcting for Galactic reddening by dividing
the sample into two subsamples, AGNs with $E(B-V) > 0.032$ and 
those with $E(B-V) < 0.032$. For these two subsamples, we calculate a
composite spectrum and find $\alpha=-0.21$ for
the high-$E(B-V)$ composite, and $\alpha=-0.50$ for the 
low-$E(B-V)$ composite.  This difference, in the sense that
the high-$E(B-V)$ composite has a harder spectrum, may indicate
that we are over-correcting for Galactic reddening.  However, the
difference is only marginally
significant, $\sim 1\sigma^{-}$ where $\sigma^{-}$ is the overall
uncertainty for the composite based 
on the total \fuse\ sample quoted above in Section~\ref{sec-uncertain}. 
The distribution in redshift and luminosity
of these two subsamples is shown in Figure~\ref{fig:lzebv}.
For the high-$E(B-V)$ subsample, the median redshift 
and luminosity are $z=0.06$ and $\log(\lambda L_{1100})= 44.5$, respectively, and
the median redshift and luminosity are $z=0.16$ and $\log(\lambda L_{1100})= 45.5$
for the low-$E(B-V)$ subsample.
Thus, the difference in spectral index between the high- and low-$E(B-V)$ subsamples
may also reflect an underlying trend toward harder EUV spectra in
lower-luminosity AGNs. 

\section{Emission Lines}
\label{sec-emlines}

\begin{figure}
\plotone{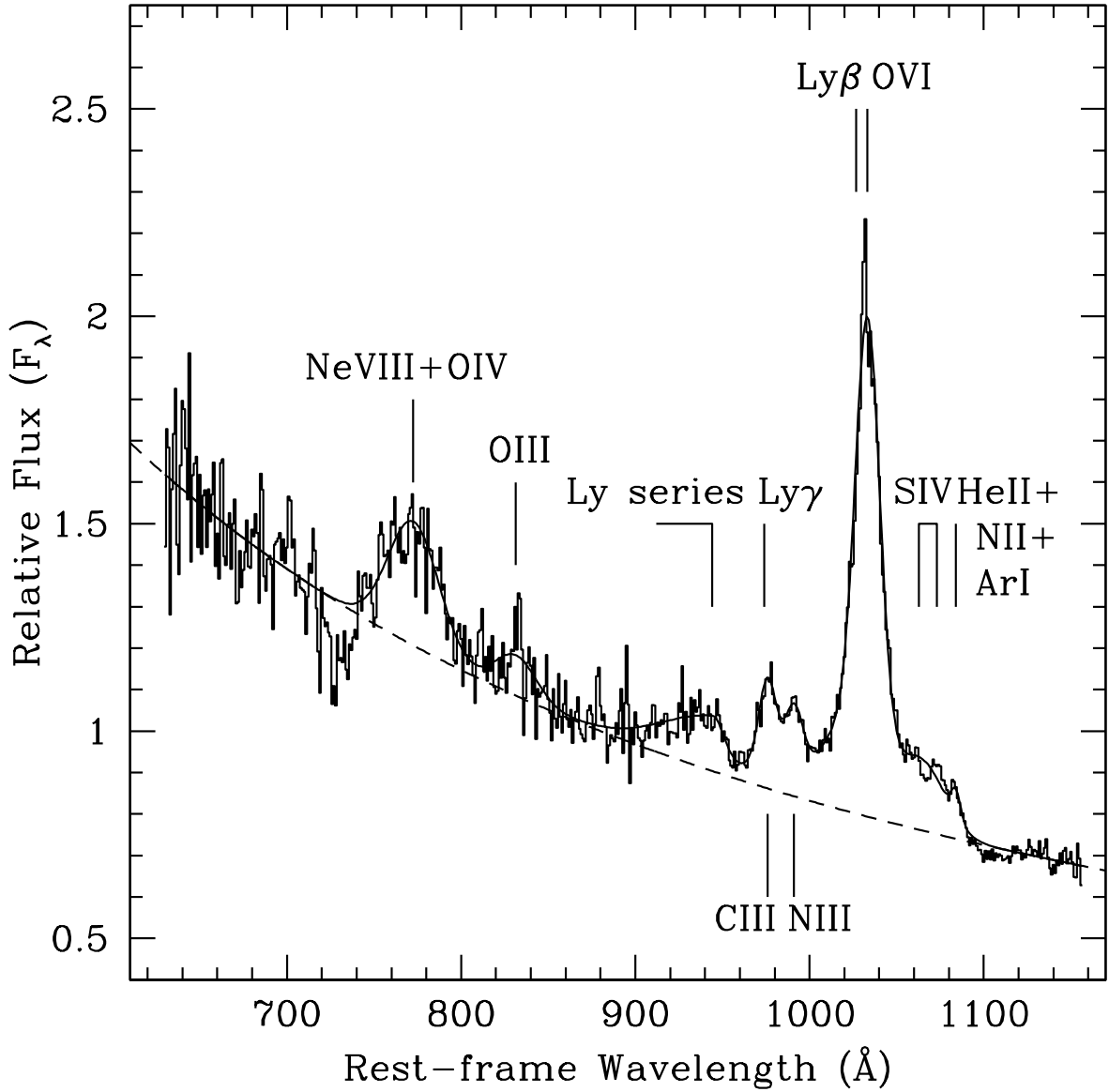}
\figcaption{Composite AGN spectrum with fits to continuum ({\it dashed line}) and continuum + 
emission lines ({\it solid line}). See Table~\ref{table-emspec} 
for fit parameters.
\label{fig:spec_em}}
\end{figure}

We fit emission lines to the \fuse\ composite spectrum 
using {\it specfit}.   We show these  lines
in Figure~\ref{fig:spec_em}, and their parameters 
are listed in Table~\ref{table-emspec}. The fluxes are normalized
such that the O~{\sc vi} emission line has a flux of 10.
We identify Ne~{\sc viii}  at 774 \AA,
O~{\sc iii} at 831 \AA, a broad feature at 944 \AA, attributed
to the \hi\ Lyman series with a possible contribution from 
S~{\sc vi}, Ly$\gamma$ at 973 \AA, C~{\sc iii} and N~{\sc iii}
at 976 and 991 \AA, respectively, Ly$\beta$ plus O~{\sc vi} at
1026 and 1033 \AA, a S~{\sc iv} doublet at 1062 and 1073 \AA,
and a feature at 1084 \AA, possibly due to N{\sc ii}, He{\sc ii}, and/or Ar{\sc i}
(Z97; T02).
A dip is present in the \fuse\ composite spectrum blueward of the 
Ne~{\sc viii} line at $\sim$730 \AA. 
This feature is visible also in the \hst\ composite from T02 in 
Figure~\ref{fig:comp}.

The bottom panel of Figure~\ref{fig:comp} shows that 
Ly$\beta$ + O~{\sc vi} and Ne~{\sc viii} are strongly
enhanced relative to the \hst\ sample.  The equivalent widths 
of these lines are $\sim$70\% and $\sim$90\% larger in the \fuse\
composite.   These spectral enhancements are the best piece of evidence
that the bluer continuum of the \fuse\ composite is not due
to a systematic error in the corrections for Galactic reddening
and/or IGM absorption but rather due to real physical
differences in AGN properties. We shall discuss this further in 
Section~\ref{sec-disc}.

\section{Spectral Fits to Individual AGNs}

\begin{figure}
\plotone{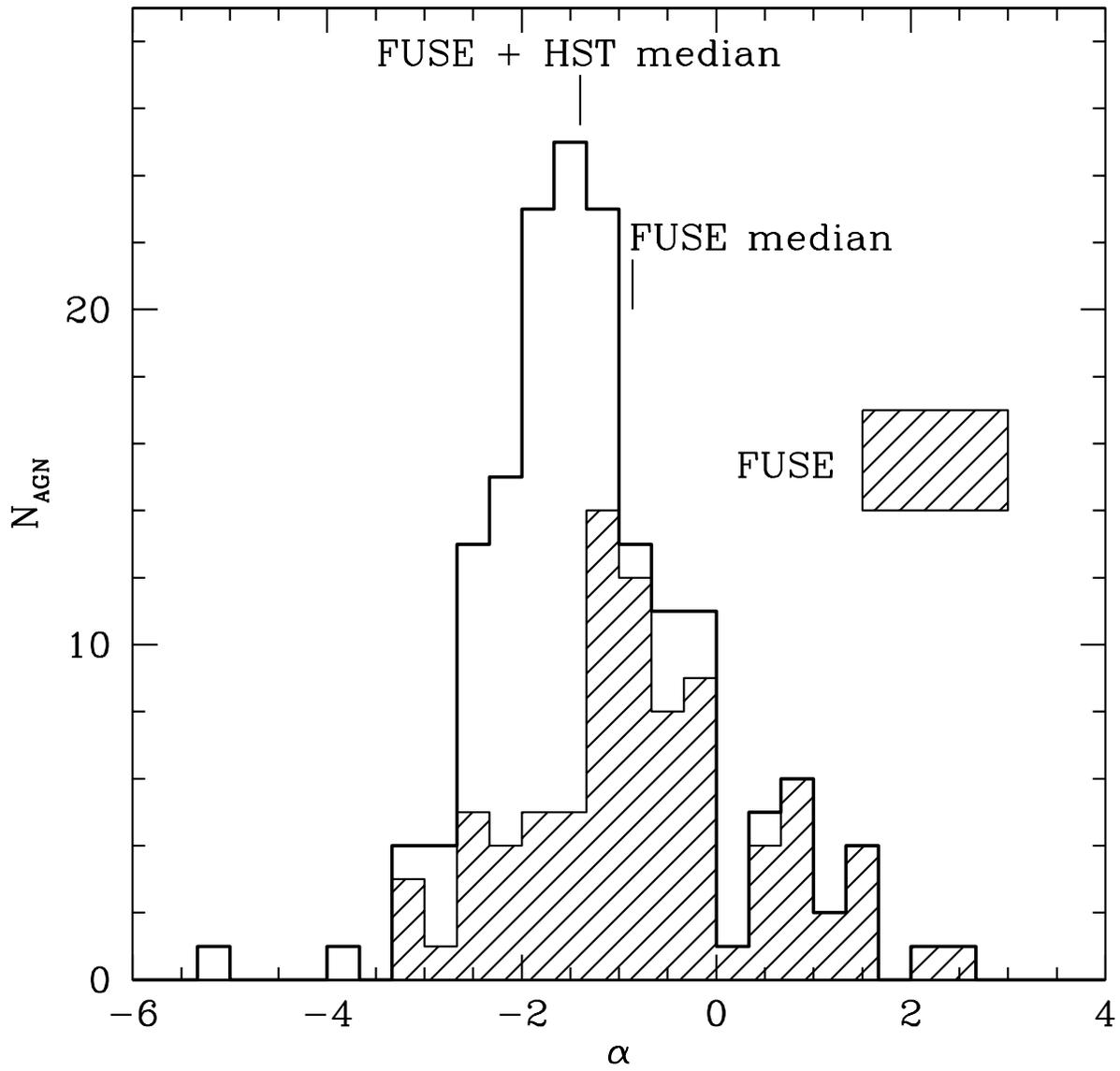}
\figcaption{
Histogram of EUV spectral slopes in \fuse\ sample and in the combined
\fuse\ and \hst\ sample.
\label{fig:ahist}}
\end{figure}

We fit a power law to each AGN EUV continuum individually for the 85 \fuse\ AGN
spectra. 
The spectral slopes, $\alpha_{\lambda}$, and extinction-corrected 
normalizations, $F_\lambda$ at 1000 \AA,
for the fits to $F_\lambda  \propto \lambda^{-\alpha_{\lambda}}$ 
are listed in columns 5 and 6 of Table~\ref{table-alpha}.
In the following discussion, we refer to the spectral slopes of the
individual AGN as defined above, $F_{\nu} \propto \nu^{\alpha}$, i.e.
$\alpha = \alpha_{\lambda}-2$.
Figure~\ref{fig:ahist} shows the distribution of the EUV
spectral slopes
in the \fuse\ sample and in the combined sample from \fuse\ and \hst,
with median values for each sample marked.  We will
discuss this combined sample further in Section~\ref{sec-hstcomp}.

\subsection{Correlations of Spectral Index with Reddening and Flux}
\begin{figure}
\epsscale{0.55}
\plotone{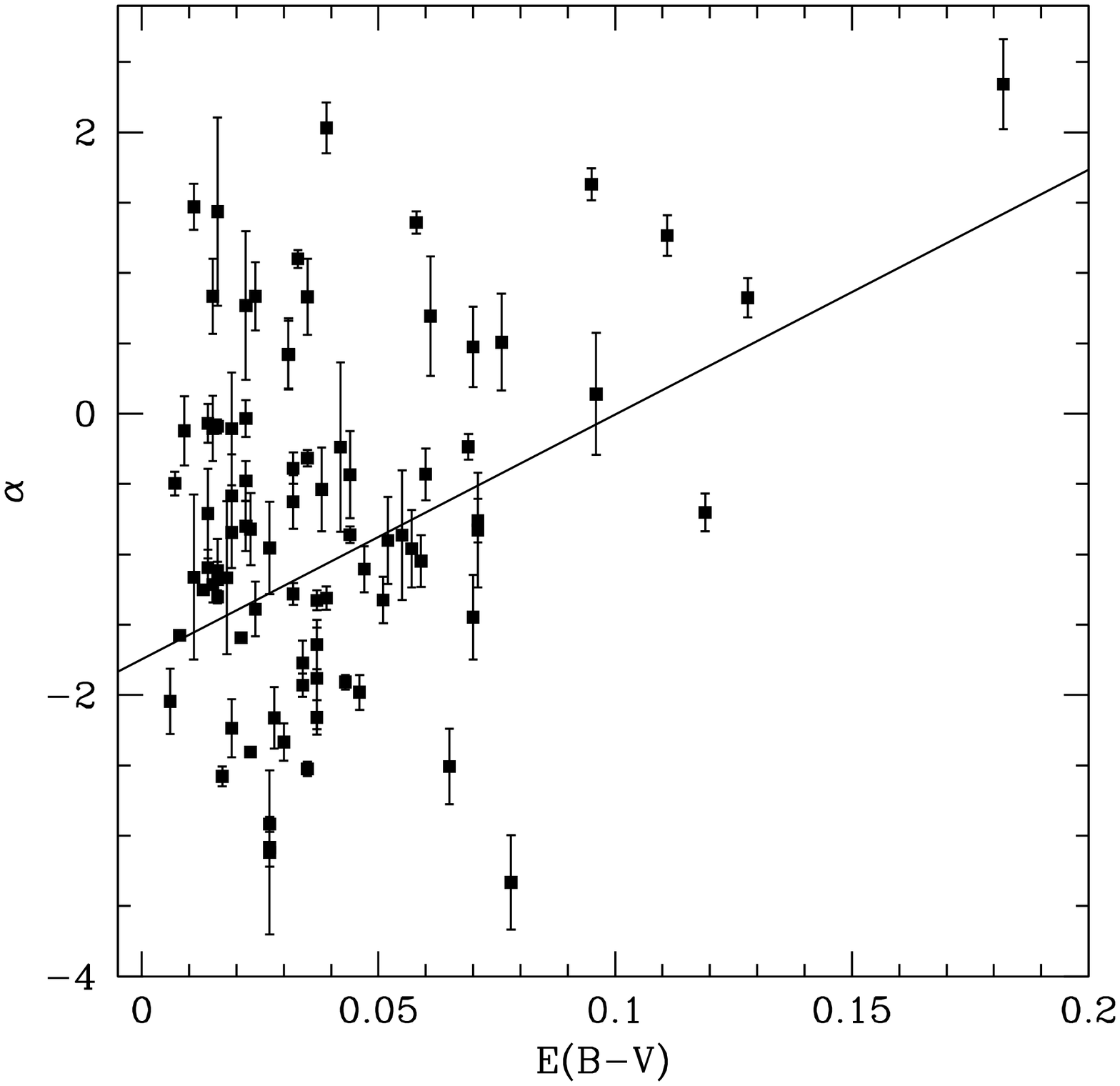}
\figcaption{
EUV spectral index versus $E(B-V)$ for FUSE AGNs with best linear least-squares fit.
\label{fig:alphaebv}}
\plotone{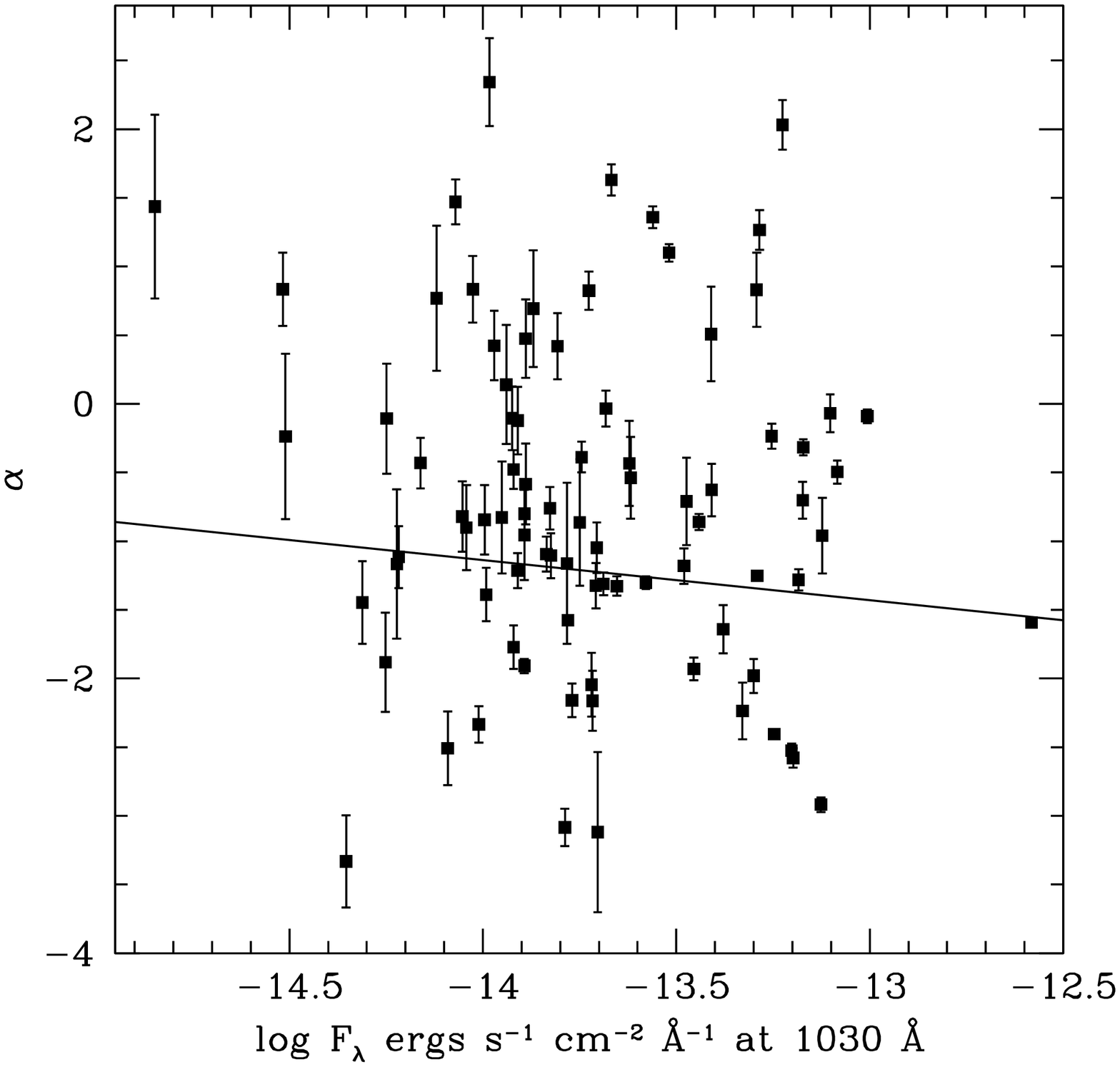}
\figcaption{
EUV spectral index versus flux at 1000 \AA\ for FUSE AGNs.
\label{fig:alphaflux}}
\epsscale{1.0}
\end{figure}

We return once more, briefly, to the issue of bias introduced by our Galactic 
reddening correction
by examining the individual
AGN spectral slopes for any trend with $E(B-V)$.  A plot of
$\alpha$ versus $E(B-V)$ 
is shown in Figure~\ref{fig:alphaebv}.
From a least-squares linear fit to these points,
we find a slope of $17.4 \pm 0.5$, but the Spearman rank-order test
indicates that the correlation is not significant (70\%).
This result gives us further confidence that our overall composite is not
significantly biased by errors in our correction of the individual
\fuse\ spectra for Galactic reddening. 

We now investigate whether our results may be influenced by systematic errors in the
background subtraction we performed on the \fuse\ data. 
This is a particular concern for faint AGNs, especially
in the short wavelength regions of the spectra, covered by the SiC channels
of the \fuse\ spectrograph, which have lower effective area than the LiF channels
by a factor of $\sim$3.  If we have systematically 
over- or under- subtracted scattered light
in the SiC channels, we expect a correlation between $\alpha$ and the observed flux.
We show the individual
spectral slopes versus the flux at 1000~\AA\ from the power law continuum fit
for each AGN in Figure~\ref{fig:alphaflux}.
The Spearman test reveals that there is no significant correlation between
these two quantities, giving us confidence that our composite does not suffer from a
bias introduced by inaccurate background subtraction in the spectra of faint AGNs.

\subsection{Correlations of Spectral Index with Redshift and Luminosity}
\begin{figure}
\epsscale{0.55}
\plotone{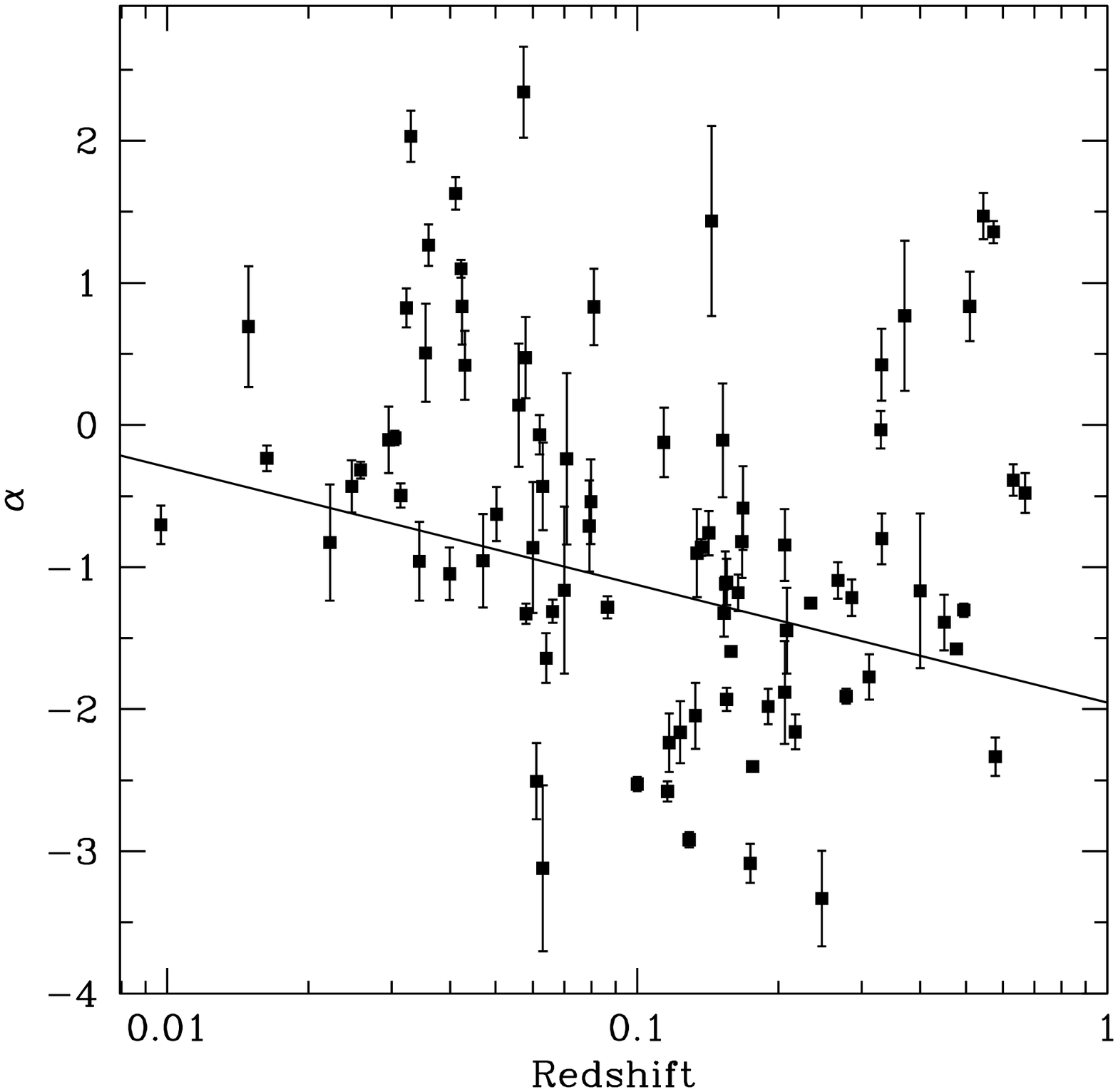}
\figcaption{EUV spectral index versus
redshift with best linear least-squares fit.
\label{fig:alphaz}}
\plotone{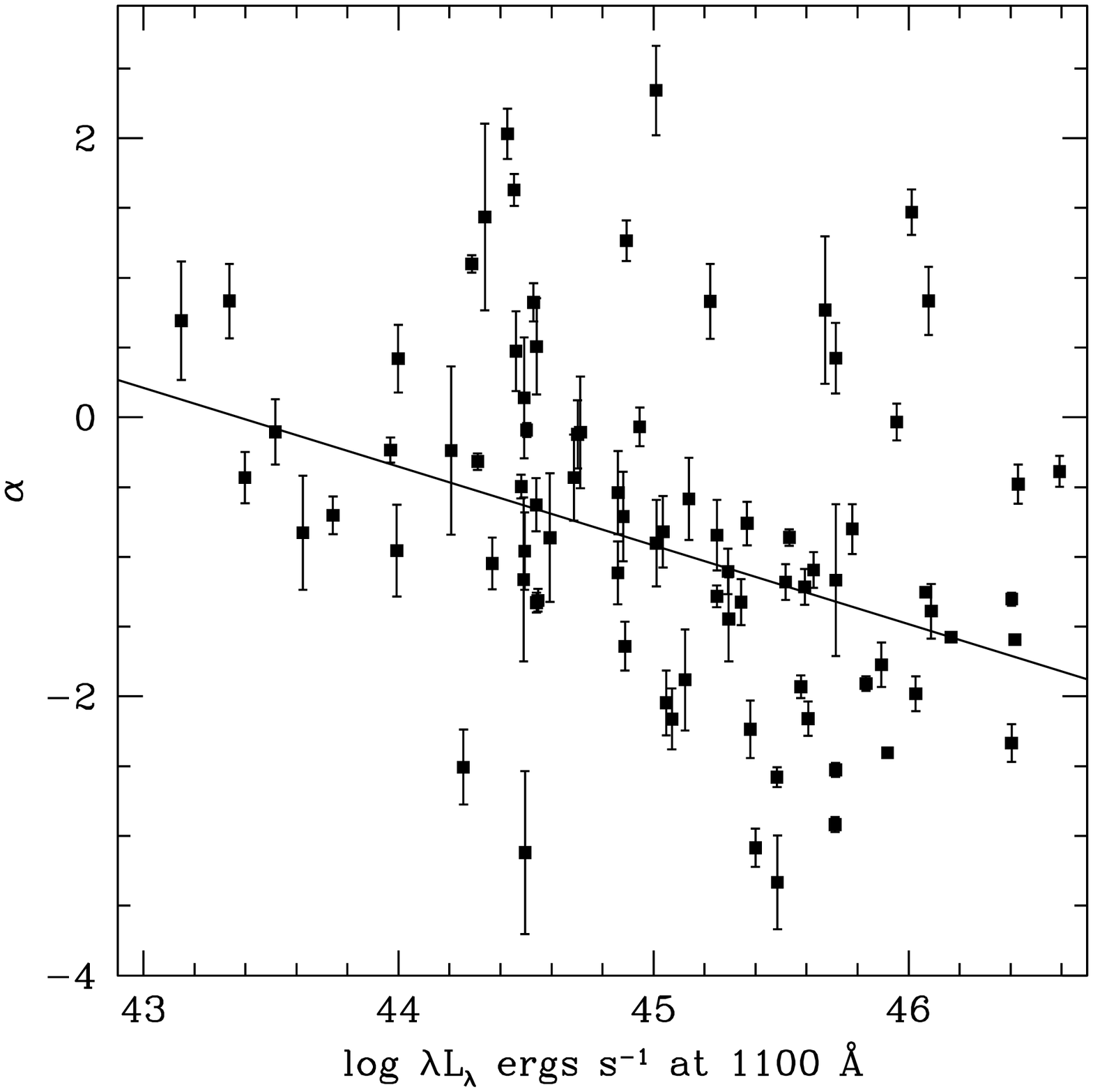}
\figcaption{
EUV spectral index versus 1100-\AA\ luminosity with best linear least-squares fit.
\label{fig:alphal}}
\epsscale{1.0}
\end{figure}

We now turn to the question of whether there are any trends 
with redshift and/or luminosity in the EUV continuum slopes
of the low-redshift AGNs in the \fuse\ sample.
We show the EUV spectral slopes versus  
redshift and luminosity in Figures~\ref{fig:alphaz} and
\ref{fig:alphal}.   
There is a significant trend of decreasing
$\alpha$ with increasing AGN redshift and luminosity in the \fuse\ sample.
The slopes of linear least-squares fits to 
$\alpha$ versus $\log z$ and $\alpha$ versus $\log(\lambda L_{1100})$
are listed in Table~\ref{table-linestats}.  The Spearman rank-order
correlation coefficients, r$_{\rm s}$, also listed in Table~\ref{table-linestats},
indicate that the correlations are significant at $>99$\% confidence. 

To test if these correlations are driven by outlier
AGNs with the hardest or softest spectral slopes,
we recalculated the correlation coefficients
excluding the 5 AGNs with the softest continua 
from the sample. We repeated this twice more, first
excluding the 10 reddest AGNs,
and then excluding the 15 reddest AGNs.  
We performed the same experiment with 
the 5, 10, and 15 sample AGNs with the hardest continua.
We found that the
correlations of spectral slope with redshift and luminosity 
were robust in all cases.
Examining composite spectra constructed from these same six subsamples,
we found that although excluding the reddest AGNs does result in a
slightly harder composite continuum, and vice versa, 
the changes in the spectral slope are not significant. 

\subsection{Comparisons with IUE and HST Samples}

\subsubsection{Redshift and Luminosity Trends in a Combined Sample}
\label{sec-hstcomp}
T02 found no significant correlation of $\alpha_{EUV}$  with redshift
or luminosity in the \hst\ AGN sample.  In view of the considerable
correlation found in the \fuse\ sample, we combine this  
with the sample of EUV spectral slopes fitted by T02 
to ascertain whether the correlations hold
for the combined sample.
As discussed above, Figure~\ref{fig:ahist} shows the distribution of $\alpha_{\rm EUV}$
in the \fuse\ sample and in the combined \fuse\ and \hst\ samples.
The mean and median spectral slopes for the \fuse\ and \fuse\ + \hst\ samples
are listed in Table~\ref{table-linestats}.
We use a bootstrap resampling method to determine the errors on the
mean and median spectral slopes, as described by T02.
We find $\alpha_{\rm median}=-0.86 \pm 0.13$ for the \fuse\
sample and $\alpha_{\rm median}=-1.40 \pm 0.09$
for the combined \fuse\ + \hst\ sample.  Both values are distinctly
harder than the medians found by T02 for the radio quiet
and radio loud QSOs in the \hst\ sample,  $-1.59 \pm 0.06$ and
$-1.95 \pm 0.12$, respectively.

\begin{figure}
\epsscale{0.5}
\plotone{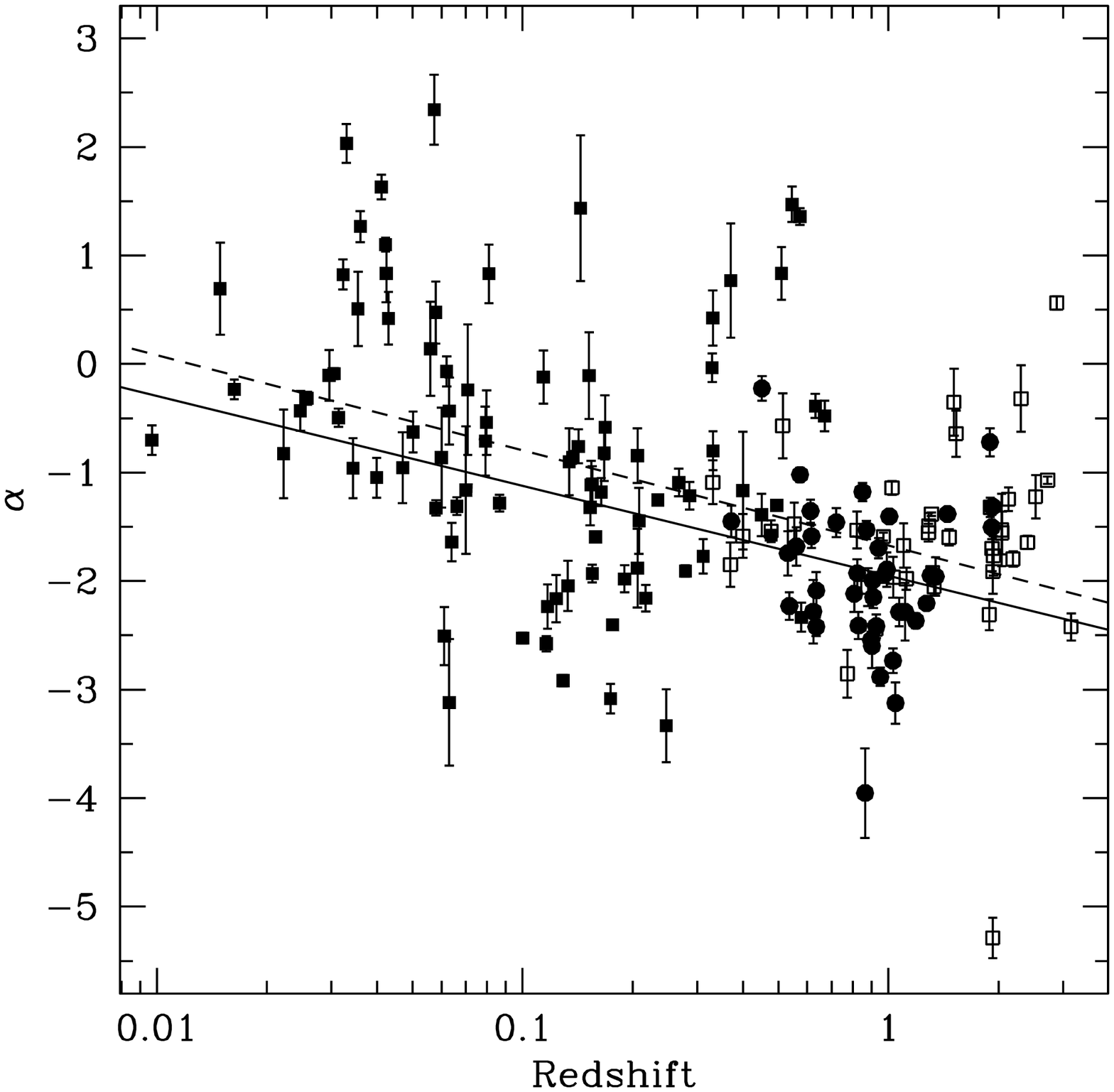}
\figcaption{Same as Figure~\ref{fig:alphaz} for both
\fuse\ ({\it solid squares}) and \hst\ (Radio-loud: solid circles, Radio-quiet: open squares)
samples with best linear least-squares fit to \fuse\ sample ({\it solid line})
and best linear least-squares fit to combined \fuse\ + \hst\ sample ({\it dashed line}).
\label{fig:alphazall}}
\plotone{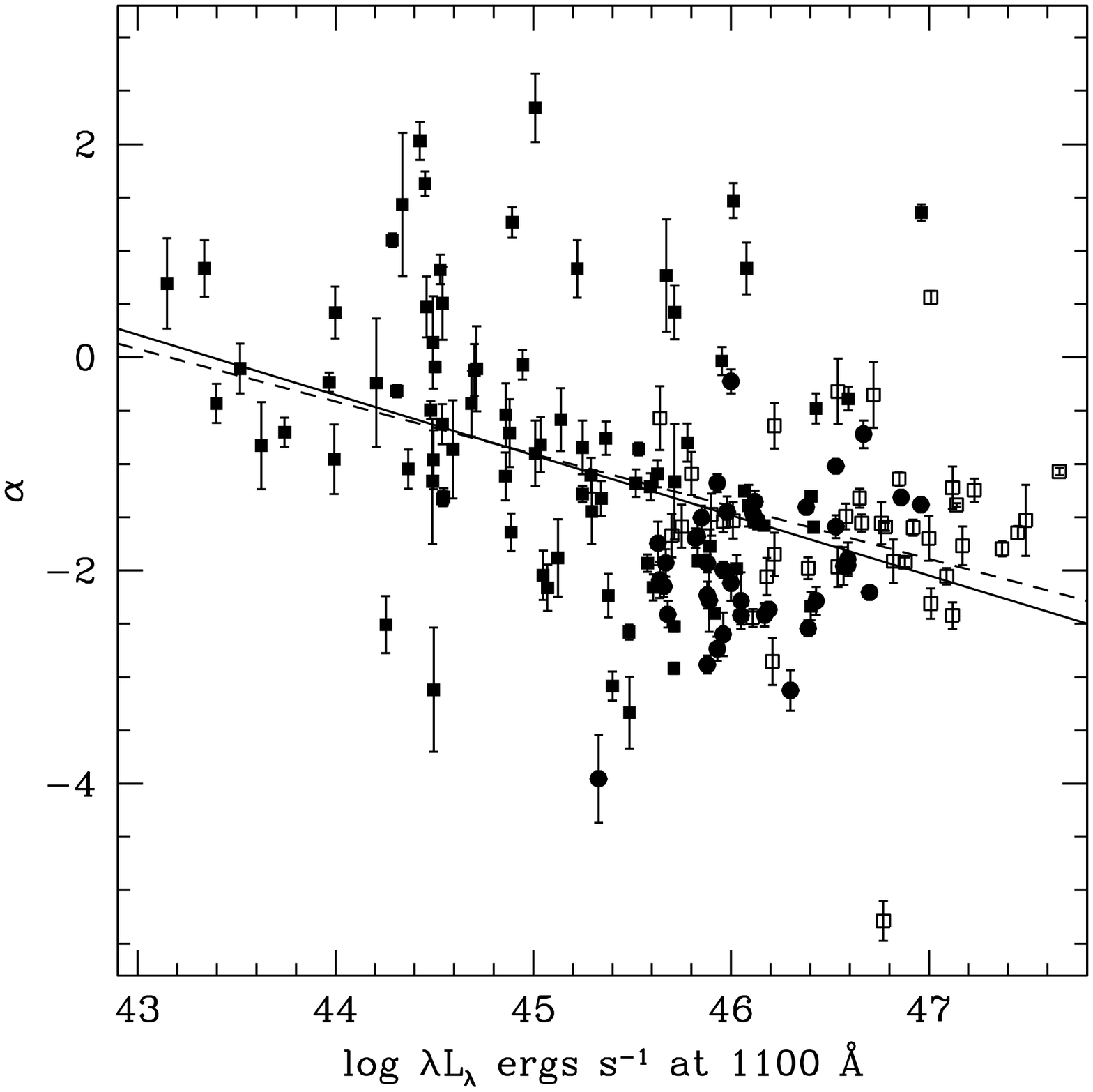}
\figcaption{
Same as Figure~\ref{fig:alphal} for both 
\fuse\ ({\it solid squares}) and \hst\ (Radio-loud: solid circles, Radio-quiet: open squares)
samples with best linear least-squares fit to \fuse\ sample ({\it solid line})
and best linear least-squares fit to combined \fuse\ + \hst\ sample ({\it dashed line}).
\label{fig:alphalall}}
\epsscale{1.0}
\end{figure}

In Figures~\ref{fig:alphazall} and \ref{fig:alphalall}, we show
the linear fits to  $\alpha$ versus $\log z$ and $\alpha$ versus 
$\log(\lambda L_{1100})$
for the \fuse\ sample and for the combined \fuse\ + \hst\ sample,
and the least-squares slopes are listed 
in Table~\ref{table-linestats}.
The correlations, in the sense
that low-redshift, low-luminosity AGNs show harder spectral slopes, are significant
at $>99$\% confidence according to the Spearman rank-order test.

\subsubsection{Near-UV/Far-UV Spectral Break}
\begin{figure}
\plotone{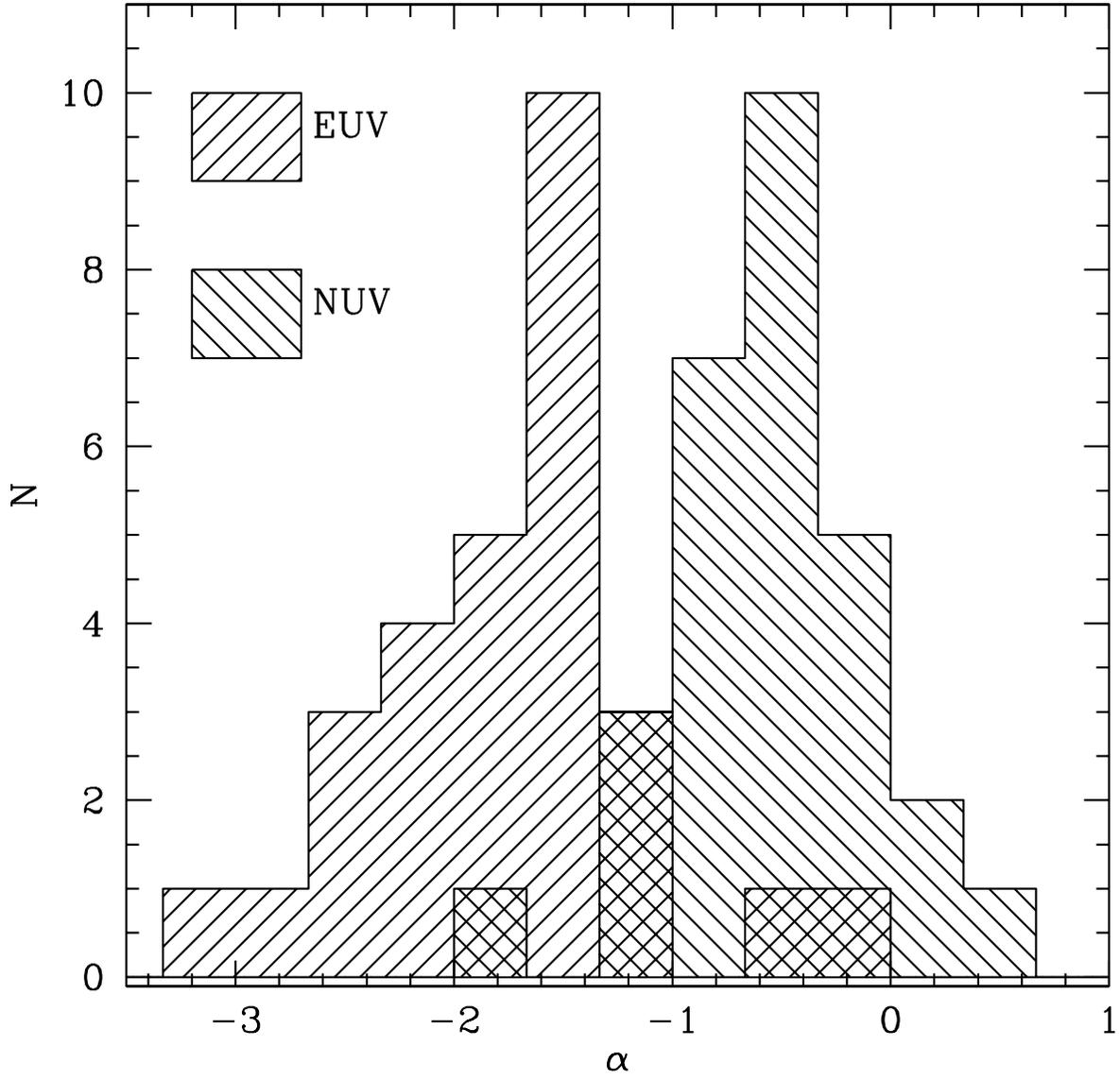}
\figcaption{
Histograms of $\alpha_{NUV}$ and $\alpha_{EUV}$ measured from \hst\ spectra of AGNs with
$z > 0.33$ by T02.  See also
their Figure 14.
\label{fig:t02hist}}
\end{figure}

T02 confirmed a break at $\sim$1200~\AA\ reported by Z97
in the spectral energy
distributions of low-redshift quasars. 
Their Figure 14 plots the individual values of 
the near-UV spectral slopes redward of $\sim$1200~\AA\ ($\alpha_{NUV}$) and the 
far-UV spectral slopes between 500~\AA\ and the break ($\alpha_{EUV}$). This
figure demonstrates that 
$\alpha_{NUV}$ is systematically
larger than $\alpha_{EUV}$.  This is also illustrated by the
histograms of $\alpha_{NUV}$ and $\alpha_{EUV}$ from T02 in Figure~\ref{fig:t02hist}.

We seek to determine whether such a break is present in the low-redshift,
low-luminosity \fuse\ AGN sample.
The \fuse\ data do not cover the near-UV, 
so we use measurements of $\alpha_{NUV}$ for our
sample AGNs from  
other datasets.
Many of the \fuse\ AGNs were observed with the {\it International
Ultraviolet Explorer (IUE)}  satellite.  
The spectral coverage of \iue, 1200-2000 \AA, is such that
the \iue\ spectral slopes, $\alpha_{IUE}$, correspond roughly to $\alpha_{NUV}$
defined by T02.
Power law continuum fits to these spectra are compiled by S.\ Penton
in an online database\footnote{See http://origins.colorado.edu/iueagn}.

We add to this sample of NUV spectral indices the 
NUV spectral slopes measured by
T02 for two objects
in common with our \fuse\ AGN sample, PG~1216+069 and PG~1543+489.
To this, we also add NUV spectral slopes measured from
combined STIS G140L and G230L spectra 
from a presently unpublished snapshot campaign to obtain
\hst\ spectra of AGNs observed with \fuse.  There are 11 of these STIS
spectra in common with the \fuse\ composite AGN sample, 8 of which
were also observed with \iue.  For the AGNs observed with both STIS
and \iue, we use the STIS spectral index for our comparison with
the \fuse\ data.  

\begin{figure}
\plotone{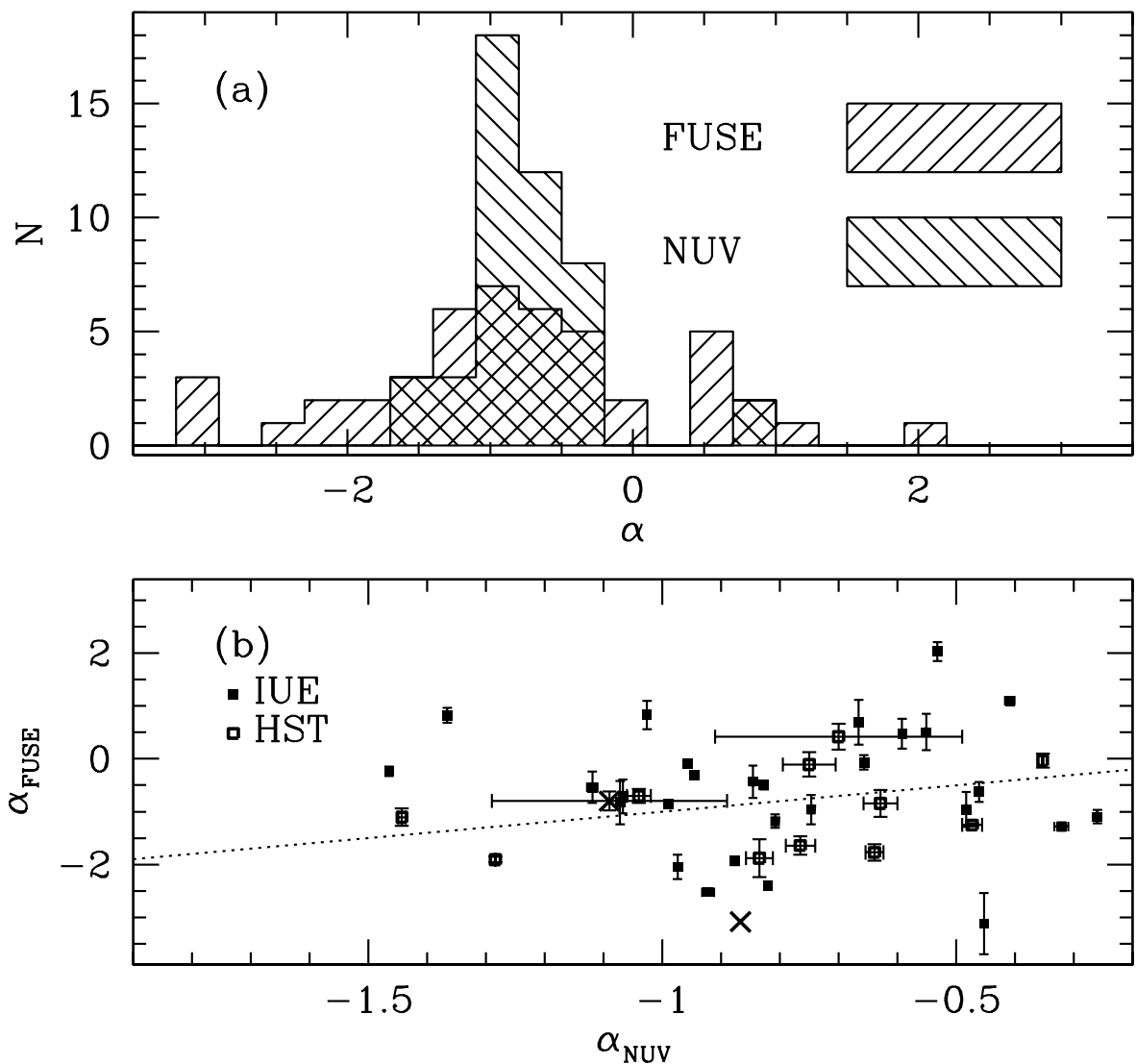}
\figcaption{
(a) Histogram of spectral slopes measured with \iue\ for rest
wavelengths 1200-2000~\AA\ or \hst\ for rest
wavelengths 1150-3200~\AA\  ($\alpha_{NUV}$) and those
measured with \fuse\ for $900 < \lambda < 1200$~\AA\ ($\alpha_{FUSE}$).
(b) Comparison of $\alpha_{FUSE}$ and $\alpha_{NUV}$.  Filled squares
denote $\alpha_{NUV}$'s measured from \iue\ spectra, open squares
denote $\alpha_{NUV}$'s measured from \hst\ spectra, and
bold crosses mark the two AGNs shown in Figure~\ref{fig:hstfuse}.
Dotted line shows $\alpha_{FUSE}=\alpha_{NUV}$.
\label{fig:alphaiue}}
\end{figure}

In Figure~\ref{fig:alphaiue}(a), 
we show histograms of the spectral slopes in the \fuse\
and \iue/\hst\ bands for the AGNs in our comparison sample. 
T02 found
$\alpha_{NUV}=-0.69$ versus $\alpha_{EUV}=-1.76$ for the full
composite \hst\ spectrum.
The median NUV and EUV spectral slopes of this AGN sample
from \iue/\hst\  and from \fuse\
are $-0.83 \pm 0.04$ and $-0.84 \pm 0.12$, respectively.
In Figure~\ref{fig:alphaiue}(b), we plot the spectral slopes
we measured from the \fuse\ spectra ($\alpha_{FUSE}$)
versus $\alpha_{HST}$ or $\alpha_{IUE}$, which
correspond to $\alpha_{NUV}$, to illustrate
$\alpha_{EUV}$ versus $\alpha_{NUV}$ in this sample, with a line denoting
$\alpha_{FUSE}=\alpha_{NUV}$.  
The distribution of the spectral slopes about this line
and the similar median EUV and NUV spectral slopes indicate that
the UV spectral break observed in the \hst\ composite is
less pronounced or even absent in the low-redshift, low-luminosity \fuse\ AGN sample.  

\begin{figure}
\plotone{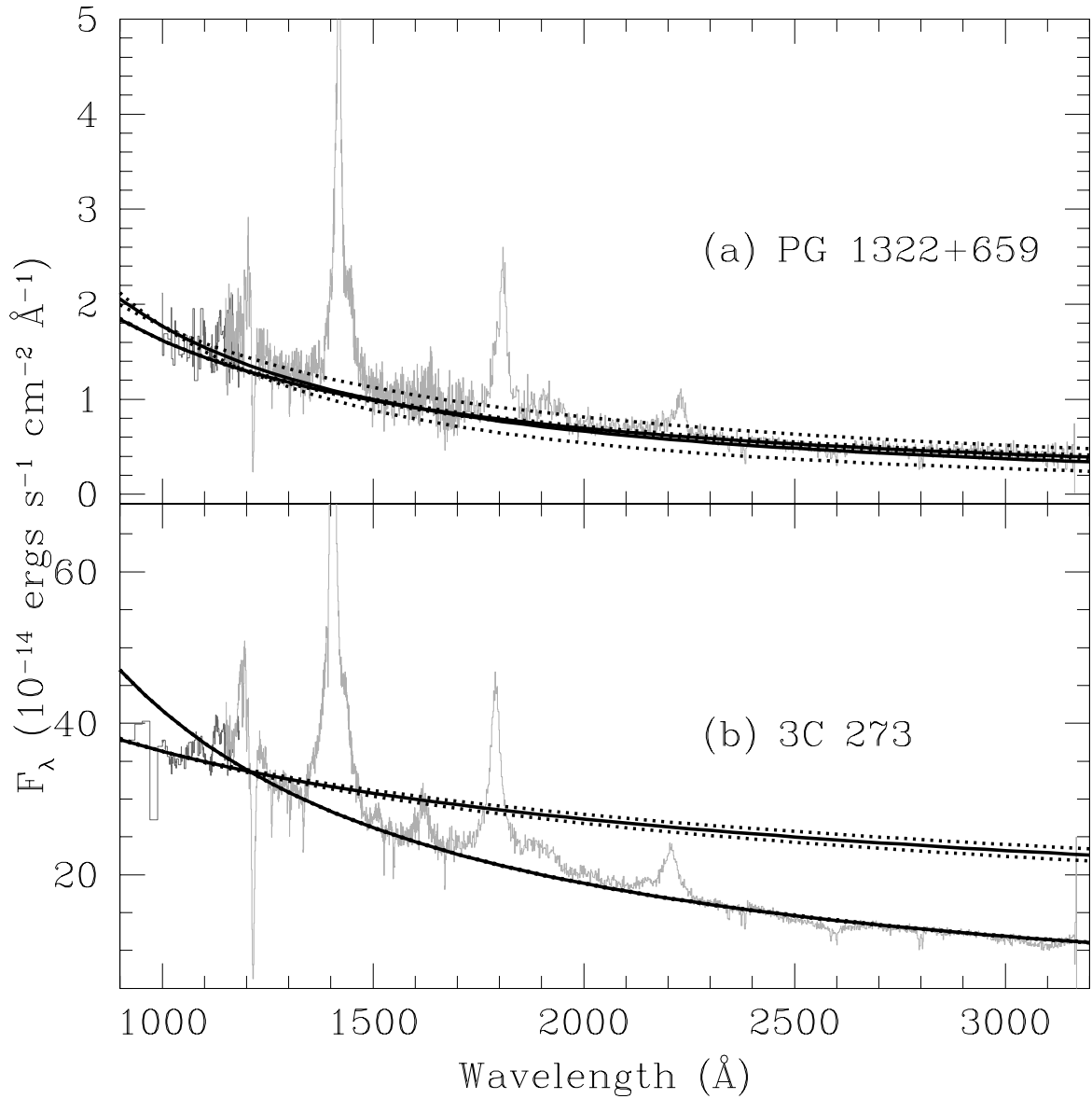}
\figcaption{
Comparison of \hst\ spectra ({\it light grey}) and smoothed
\fuse\ spectra ({\it dark grey}) of
PG~1322+659 ({\it top panel}) and 3C~273 ({\it bottom panel}).
The solid lines show the power law continuum fits, and the
dotted lines indicate the 1$\sigma$ uncertainties on the
power law index.
\label{fig:hstfuse}}
\end{figure}

While the \fuse\ AGN do not appear to show a spectral break as an ensemble,
some individual objects in the sample do show a break in the far-UV.
In Figure~\ref{fig:hstfuse},
we show two examples of AGNs with both \fuse\ and STIS G130L+G240L spectra,
one with no break between the STIS and \fuse\ bands, and one with a distinct break. 
These two examples are marked as bold crosses
in Figure~\ref{fig:alphaiue}(b).  
In the top panel, we show the \fuse\ and STIS spectra of
PG~1322+659, a Seyfert 1 galaxy with $z=0.168$.  The UV spectrum of this AGN
shows no evidence for a break, $\alpha_{EUV}=-0.58 \pm 0.29$ and
$\alpha_{NUV}=-0.78 \pm 0.04$.
Some low redshift AGN do show a spectral break, however.  Kriss et al.\ (1999) reported
a break in the spectrum of 3C~273 at $\sim$900~\AA\ from {\it Hopkins
Ultraviolet Telescope} data, and they
interpreted that break 
as the signature of a Comptonized accretion disk spectrum (Shields 1978; 
Malkan \& Sargent 1982).
Our comparison of the \fuse\ and
STIS spectra of this object, shown in Figure~\ref{fig:hstfuse}(b),
confirm a significant break between the NUV and EUV spectral slopes:
$\alpha_{EUV}=-1.60 \pm 0.03$ and $\alpha_{NUV}=-0.856 \pm 0.005$. 
Note that we simply confirm a spectral break and 
have not attempted a self-consistent double power law
fit to the full UV continuum here.
A thorough study of the EUV-to-optical spectral energy distributions of
AGNs and comparisons with accretion disk models
using the STIS snapshot data reported above 
is forthcoming (Shang et al., in preparation).

\subsection{Correlation of Spectral Index with Accretion Properties}
\begin{figure}
\epsscale{0.55}
\plotone{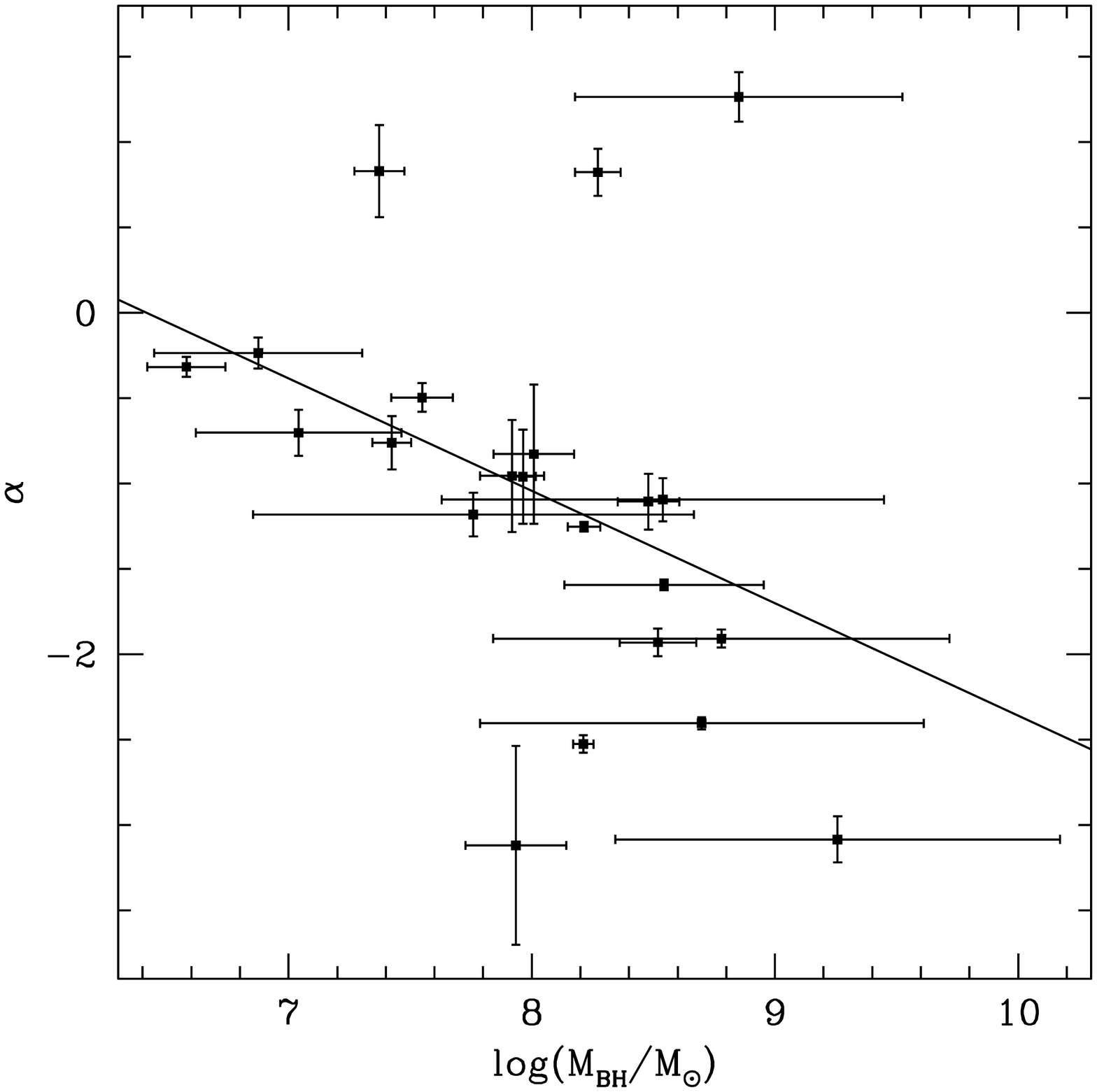}
\figcaption{
EUV spectral index versus black hole mass listed in Table~\ref{table-alpha} 
for 21 AGNs in the \fuse\ sample, with best linear least-squares fit.
\label{fig:alphambh}}
\plotone{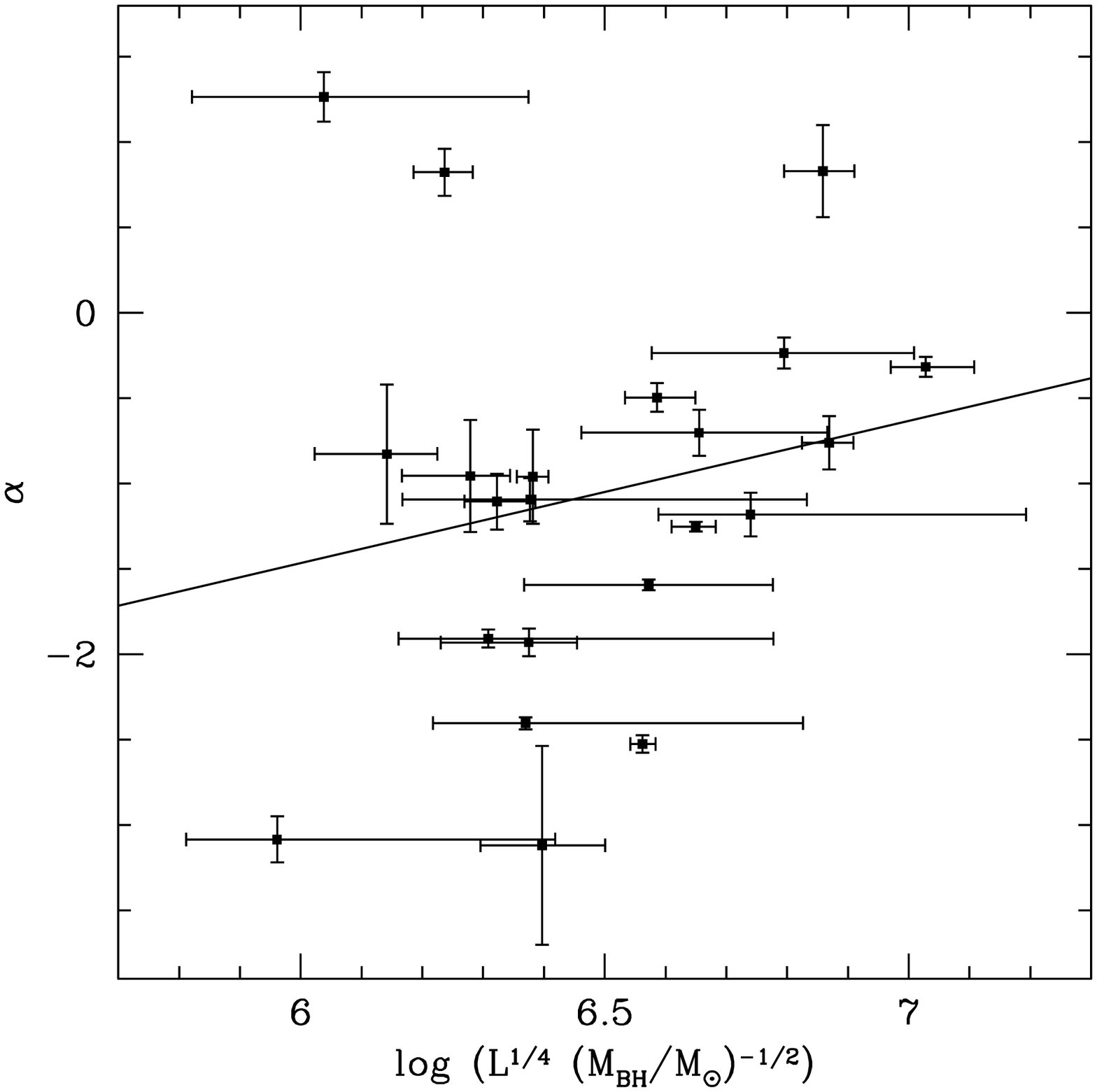}
\figcaption{
EUV spectral index versus temperature ($\sim \log( L^{1/4} M_{BH}^{-1/2} )$)
with best linear least-squares fit.
\label{fig:alphambh2}}
\epsscale{1.0}
\end{figure}

Here we investigate whether the spectral slopes of the low-redshift
AGNs in the \fuse\ sample are correlated with observable properties of
the central black hole or the mass accretion process.
For 22 AGNs in the \fuse\ sample, we 
have estimates of black hole mass from Kaspi et al.\ (2000) and 
M$^{\rm c}$Lure \& Dunlop (2001).
This quantity is listed in column 6 of Table~\ref{table-alpha} and is plotted
against the spectral indices of individual AGNs in Figure~\ref{fig:alphambh}.
We find a correlation of spectral slope with
black hole mass, at 96\% confidence according to the
Spearman rank-order test.  The trend runs in the sense that AGNs with lower black
hole masses show harder spectral slopes.  A linear least-squares fit
gives a slope of $-0.67 \pm 0.35$, 
shown by the line in Figure~\ref{fig:alphambh}.

For standard, geometrically thin, optically thick accretion disks, 
the temperature of the disk is proportional to its luminosity and to 
the mass of the central black hole as:
$T \sim L^{1/4} M_{BH}^{-1/2}$ (Shakura \& Sunyaev 1973).  Therefore, 
the thermal disk emission will peak further
in the blue in the spectra of AGNs with lower mass black holes.
Thus, the spectral index observed in the NUV will extend
through the EUV, and we may expect a correlation between
$\alpha$ and temperature.
We show $\alpha$ versus $L^{1/4} M_{BH}^{-1/2}$
in Figure~\ref{fig:alphambh2} and the linear least-squares fit
with a slope $0.86 \pm 0.91$.  The slope has the expected sign, but the
correlation is not statistically significant, 70\%.

\section{Discussion}
\label{sec-disc}

The \fuse\ AGN sample is approximately one-half the size of the
\hst\ sample presented by T02, and it is distinctly different in terms of
both redshift ($z_{\rm median}=0.10$ versus $z_{\rm median}=0.96$ for the
\hst\ AGNs) and luminosity (median $\log(\lambda L_{1100})=45.0$ versus
median $\log(\lambda L_{1100})=45.9$).  
Like the Z97 and T02 \hst\ samples, this
\fuse\ archival sample is a heterogeneous sample of AGNs observed with
\fuse\ for various reasons.  Presumably, however,
most AGNs in the sample were observed because they were known {\it a priori}
to be bright in the UV, particularly in the NUV. However, we note that
Figures~\ref{fig:lz}
and \ref{fig:hist} illustrate that the sample is well-populated with
intrinsically low-luminosity AGNs.
The EUV spectral index of the \fuse\ composite is significantly
harder than the \hst\ composite, $\alpha= -0.56^{+0.38}_{-0.28}$
versus $\alpha=-1.76\pm0.12$.

In constructing our composite spectrum,  we
followed the methodology of Z97 and T02 by using uniform
weighting for each AGN spectrum.   This method prevents the
overall composite spectrum from being dominated by the 
brightest AGN with spectra of the highest S/N, such
as 3C273 ($\alpha=-1.60$).
This in turn provides the best estimate
of the spectrum of the average AGN (Z97) and the best comparison
with the \hst\ results of Z97 and T02.

At face value, the harder spectral slope of a composite 
formed from low-redshift AGNs seems to lend credence to 
a scenario in which the spectral break in the \hst\
composite spectrum reported by Z97 and T02 is caused by
an undercorrection for intergalactic absorption, as proposed 
by Binette et al.\ (2003).  However, 
this scenario does not easily explain the correlation
between spectral slope and AGN luminosity we report, nor
does it account for the enhanced emission lines in the \fuse\
composite relative to the \hst\ composite.  

A more natural explanation of these results is that we are
seeing a manifestation of the physical mechanism put forth to
explain the Baldwin effect. 
This correlation is generally attributed to the tendency for
low-luminosity AGNs tend to show harder ionizing continua
(Zheng \& Malkan 1993; Wang et al.\ 1998; Dietrich et al.\ 2002).
This suggests that 
both the enhanced high-ionization emission line strengths and
the harder continuum shape of the \fuse\ composite spectrum
with respect to the \hst\ composite
are due to the larger fraction
of relatively low-luminosity AGNs in the \fuse\ sample.  
T02 explain the excess of C~{\sc iv} emission in the \hst\
composite relative to the SDSS composite (Vanden Berk et al.\ 2001)
in the same way.
The  larger equivalent widths of the high-ionization emission
lines O~{\sc vi} and Ne~{\sc viii} in the \fuse\ composite
relative to the \hst\ composite is the best evidence that the
harder spectral index of the \fuse\ composite is not caused by
any wavelength-dependent systematic error in our corrections 
for Galactic and intergalactic absorption.
In addition, the significant correlation between individual AGN spectral
slopes and luminosities for both the \fuse\ sample and the
combined \fuse\ + \hst\ samples supports the Baldwin effect
interpretation.

\begin{figure}
\plotone{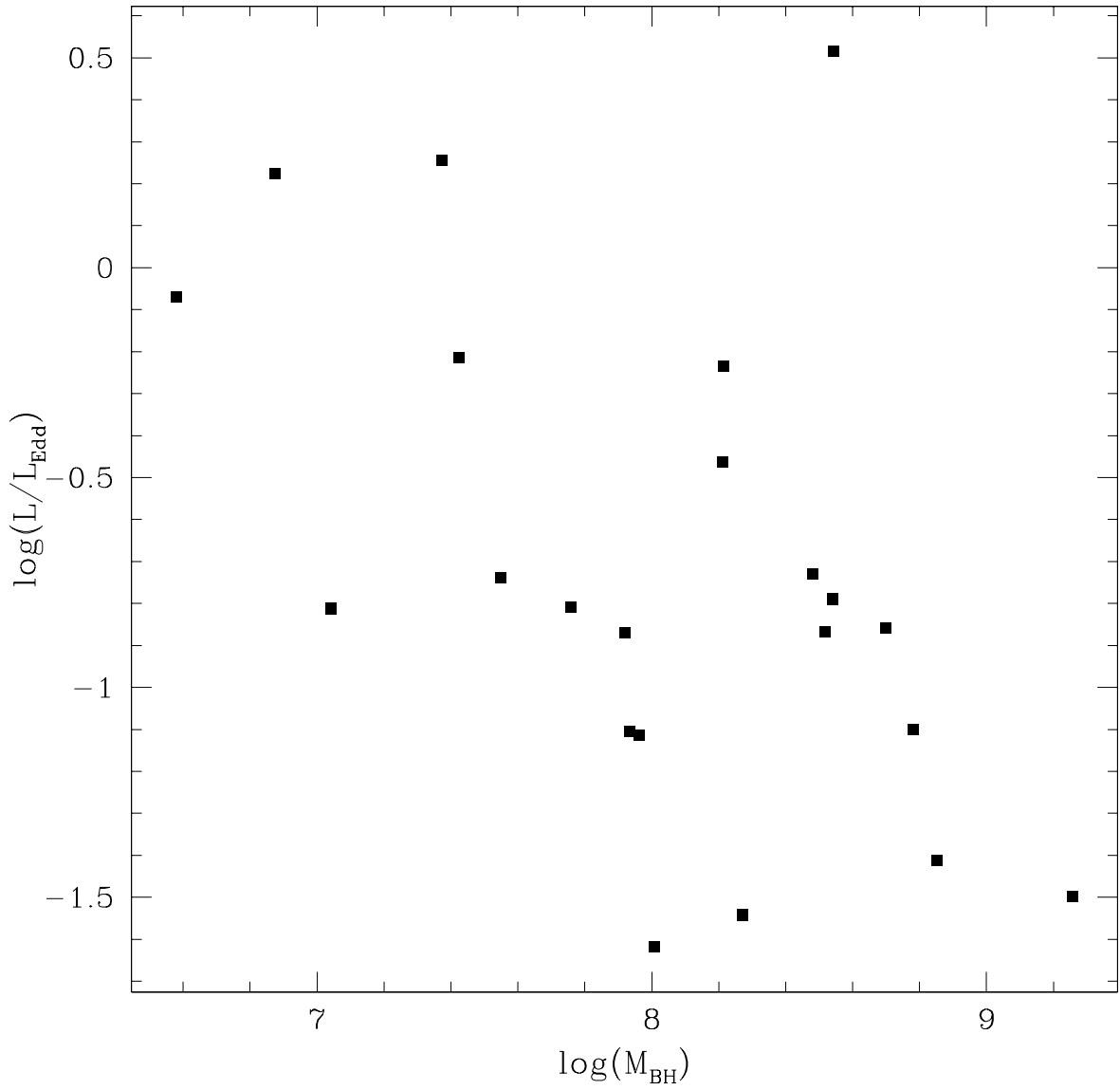}
\figcaption{Eddington ratio versus black hole mass for 20 \fuse\ AGNs.
See Table~\ref{table-alpha}.
\label{fig:ledd}}
\end{figure}

Z97 interpreted the spectral break at 1050~\AA\ in their \hst\ 
composite, along with a depression of the flux near the Lyman limit,
as the signature of a Comptonized accretion disk spectrum.
Relativistic effects on the emergent spectrum from an inclined accretion disk
as well as Comptonization from a hot corona can broaden an intrinsic Lyman
limit break (Lee, Kriss, \& Davidsen 1992).
Z97 estimated a Lyman limit optical depth $\tau=0.8$ for a 10\% depression in the flux.
We have shown that the AGNs in this low-redshift, low-luminosity \fuse\ sample
tend to lack such a spectral break, and we find no evidence for a depression
in the spectrum blueward of the Lyman limit in the \fuse\ composite.  
From the S/N in the composite,
we estimate that any spectral discontinuity is less than 10\%.
Using non-LTE accretion disk models
with no Compton scattering,
Hubeny et al.\ (2000) demonstrated that emission may wash out 
Lyman edge discontinuities in QSOs with
$L \gtrsim 0.01 L_{\rm Edd}$, particularly
in systems with $M_{\rm BH} \lesssim 10^{9} M_{\sun}$.
We calculate the Eddington ratio for the 22 \fuse\ AGNs for which we have
estimates of black hole mass in Table~\ref{table-alpha}
using that black hole mass and bolometric luminosities from Woo \& Urry (2002)
or Padovani \& Rafanelli (1988), also listed in Table~\ref{table-alpha}.
The Eddington ratio is plotted versus black hole mass in Figure~\ref{fig:ledd}, which
indicates that the \fuse\ AGN occupy the 
region of parameter space noted above.  The lack of a flux
discontinuity in the \fuse\ composite is consistent with
the Hubeny et al.\ (2000) models in this sense. However, their models
do predict a change in the spectral slope for these AGNs at wavelengths
blueward of the Lyman limit, a break we do not find in the \fuse\ composite.

The EUV spectral slope found here for AGNs with $z \le 0.67$ 
does imply a spectral break between
the far UV and the soft X-rays given the soft X-ray slopes
reported by Laor et al.\ (1997), who find $<\alpha_{x}>= -1.72\pm0.09$
and $<\alpha_{x}>= -1.15\pm0.14$ for radio-quiet and radio-loud low-redshift 
QSOs, respectively.
It is consistent with the presence of a soft X-ray excess and steep
soft X-ray power law to join with the 0.2-2~keV power law, as 
Mathews \& Ferland (1987) inferred from the observed strength of 
He~{\sc ii}~$\lambda$1640 emission.  The hard slope of the \fuse\
composite may also help resolve questions raised by Korista, Ferland, \&
Baldwin (1997) about how a soft EUV continuum such as the one
found by Z97 and T02 could account for the observed strength of this line.
Neglecting any turnover in the EUV-to-soft-X-ray spectral energy distribution, the
\fuse\ composite predicts about eight times as many photons
at 4 Ryd as the $\alpha=-2$ power law these authors used to calculate
He~{\sc ii}~$\lambda$1640 equivalent widths of 0.6-0.8~\AA\ for 
covering fractions of 10\%.  The \fuse\ data do not cover rest frame 
1640~\AA, and although
He~{\sc ii}~$\lambda$1085 emission line may be present in
the emission line complex identified in the composite at 1084~\AA\ 
(see \S~\ref{sec-emlines}), this identification is uncertain.
We therefore leave a more complete 
consideration of the implications of the \fuse\ EUV spectral
slope on AGN emission lines for future work.
 
If AGNs build up black hole mass via accretion
over their lifetimes,  the evolutionary models of Wandel (1999a,b) predict
higher mass (luminosity) AGNs will have lower temperature accretion disks and
softer spectral slopes in the UV and X-rays.
The absence of a spectral break in the \fuse\ composite
can be explained by such models.
Low-luminosity
AGNs possess hotter accretion disks, and this shifts the UV bump and the Compton break
to shorter wavelengths (Z97).
The anticorrelation we find between EUV spectral slope
and black hole mass estimates compiled from the literature also supports this 
interpretation.  This
is undermined somewhat by the low significance of the correlation between
EUV spectral index and accretion disk temperature, estimated by $L^{1/4} M_{BH}^{-1/2}$.
For AGNs accreting at the Eddington limit, 
$L \propto M_{BH}$ and $T \propto M_{BH}^{-1/4}$ and
the significant correlation between $\alpha$ and $M_{BH}$ may directly reflect a trend of
$\alpha$ and accretion disk temperature.
The Eddington ratios listed in Table~\ref{table-alpha} and plotted in
Figure~\ref{fig:ledd} indicate
that in fact most of the AGNs in the sample shine at luminosities below the Eddington limit.
However, we note that the estimate we use for the accretion disk temperature
is a combination of two observables.
If a trend between spectral slope and disk temperature 
is indeed present, it could easily be swamped by the considerable
uncertainties in the luminosity and black hole mass.

Finally, we discuss the dip in the composite spectrum blueward of the
Ne~{\sc viii} emission line.  Possibly, this arises from the superposition of 
absorption features arising from highly ionized gas along the line of sight, either
intrinsic to the AGNs or in the IGM.
It is unlikely to reside in intergalactic absorbers comprising the warm-hot IGM (WHIM).
These absorbers have been observed in O~{\sc vi} in
several QSO sightlines
(Tripp, Savage, \& Jenkins 2000; Oegerle et al.\ 2000; 
Tripp \& Savage 2000; Tripp et al.\ 2001; Sembach et al.\ 2001; 
Savage et al.\ 2002).  
We see no corresponding dip in the flux at the position of the O~{\sc vi} doublet
in the \fuse\ composite spectrum.
Whether these absorbers are photoionized or collisionally ionized, 
it is likely that
O~{\sc vi}/Ne~{\sc viii} $ \gg 1$ (Tripp \& Savage 2000). 
On the other hand,
if the absorbing gas is collisionally ionized and 
$5.75 \lesssim \log T \lesssim 6.00$, or if nonequilibrium conditions hold, 
the abundance of Ne~{\sc viii} 
could be appreciable, even comparable to that of O~{\sc vi} (Heckman et al.\ 2002).
At such temperatures, O~{\sc vii} and even O~{\sc viii} absorption
would be expected and have been observed (Nicastro et al.\ 2002; Mathur, Weinberg, \& Chen 2003).
Nonetheless, if this flux depression is attributable to Ne~{\sc viii} in the WHIM,
the question of why we find no depression due to intervening O~{\sc vi} is still open.
If we do attribute the dip to intervening
Ne~{\sc viii} absorption, we may
estimate the mean cosmological mass density in these
absorbers using
\begin{eqnarray}
\Omega_{b}( \mathrm{Ne}~\mathrm{\scriptstyle VIII} ) =
\frac{\mu m_{H}H_{0}}{ \rho_{c} c f(\mathrm{Ne}~
\mathrm{\scriptstyle VIII}) }
  \left( \frac{\mathrm{Ne}}{\mathrm{H}} \right)^{-1}  \nonumber \\
 \times \frac{\sum_{i} N_{i}(\mathrm{Ne}~\mathrm{\scriptstyle VIII} ) }{ \Delta X }
\end{eqnarray}
(Tripp et al.\ 2000),
where $\mu$ is the mean atomic weight, taken to be 1.3, 
$f$(Ne~{\sc viii}) is the ionization fraction of Ne~{\sc viii}, and
(Ne/H) is the neon abundance by number.
The quantity $\sum_{i} N_{i}(\mathrm{Ne}~\mathrm{\scriptstyle VIII} )$
is the total column density of Ne~{\sc viii}
estimated from the apparent column density  in the
absorption dip in the composite.  The effective optical depth of
the feature implies a total column density $N(\mathrm{Ne}~\mathrm{\scriptstyle VIII} )
 \sim 6 \times 10^{15} {\rm cm}^{2}$,
assuming optically thin absorption (Savage \& Sembach 1991).
The term $\Delta X$ is the total absorption distance probed by all
the sample AGN assuming $q_{0}=0.5$ (Bahcall \& Peebles 1969).
For [Ne/H]=-1 and $f$(Ne~{\sc viii})=0.2, the peak
fractional abundance of Ne~{\sc viii} at $\log T =5.85$ (Shapiro \& Moore 1976),
we estimate $\Omega_{b} \sim 0.027 h_{75}^{-1}$.
This value is 
a factor of $\sim$10 larger than the more careful lower limits set 
from O~{\sc vi} absorbers (Tripp \& Savage 2000; Savage et al.\ 2002), and
it would imply that over 60\% of the baryons at $z<0.7$
reside in Ne~{\sc viii} absorbers.

This flux depression has a symmetric appearance
with a velocity extent of $\sim$10,000 km~s$^{-1}$, and 
the centroid of the absorption 
lies within $\sim$17,000 km~s$^{-1}$ of the Ne~{\sc viii} emission line.
It is difficult to imagine a redshift distribution of intervening
absorbers that would naturally give rise to these characteristics,
which are more closely akin to broad-absorption-line (BAL) systems.
It is more likely that this feature is associated with the AGNs themselves.
Broad, intrinsic highly-ionized absorption from
Ne~{\sc viii} has been observed in several high redshift QSOs
(Q~0226-1024: Korista et al.\ 1992;
SBS~1542+541: Telfer et al.\ 1998; and
PG~0946+301: Arav et al.\ 1999),
as have narrow associated absorption
features (UM~675:  Hamann et al.\ 1995, 1997;  
HS~1700+6416: Petitjean, Riediger, \& Rauch 1996; 
J2233-606: Petitjean \& Srianand 1999; and
3C~288:  Hamann, Netzer \& Shields 2000).
We find no obvious Ne~{\sc viii} absorption features in
any single AGN spectrum.  We deliberately excluded AGNs known
to be BALs from the sample, but many AGN showing narrow intrinsic UV absorption
troughs associated with X-ray $``$warm absorbers", such as 
Mrk~279 (Scott et al.\ 2004),
Mrk~509 (Kriss et al.\ 2000a; Kraemer et al.\ 2003),
NGC~3783 (Gabel et al.\ 2003), and
NGC~7469 (Kriss et al.\ 2000b; Kriss et al.\ 2003)
are included in the 
\fuse\ composite sample.
In  photoionization models of intrinsic absorbers, Ne$^{7+}$ and
O$^{6+}$ can coexist over large regions (Hamann 1997; Wang et al.\ 2000).
Furthermore, given solar abundances and
the similar oscillator strengths
of these two doublets,
we might expect to find a corresponding dip in the composite 
from associated O~{\sc vi} absorption.
However, the BAL in the spectrum of SBS~1542+541 shows absorption
due to highly-ionized species, including Ne~{\sc viii}, and lacks
absorption features from low-ionization species such as
C~{\sc iii} and Si~{\sc iv} seen in other BALs (Telfer et al.\ 1998). 
Their results
suggest that the covering fractions of various ions can
be dependent upon creation ionization potential.
They do find O~{\sc vi} absorption in this BAL system, but our selection
against known BALs in the \fuse\ sample from features like
C~{\sc iv} and  O~{\sc vi}, combined with an ionization-dependent
covering fraction is a plausible explanation of the flux depression
blueward of Ne~{\sc viii} in the \fuse\ composite, and the lack of
one blueward of O~{\sc vi}.

\section{Summary}
We summarize our results as follows:

\begin{enumerate}

\item{We construct a composite EUV (630-1155 \AA)
spectrum of AGNs with $z \le 0.67$ from archival \fuse\ data.}

\item{We find that the best-fit spectral index of the composite is
$\alpha=-0.56^{+0.38}_{-0.28}$.
The conservative estimate of the total error in the spectral index
includes the standard deviation in $\alpha$ in
1000 bootstrap samples of the \fuse\ data set,
uncertainties in the extinction correction applied
to the \fuse\ spectra, and uncertainties in the
column density distribution parameter of the intervening Ly$\alpha$ forest.}

\item{We find that O~{\sc vi}/Ly$\beta$ and Ne~{\sc viii} emission  are enhanced
in the \fuse\ composite relative to the \hst\ composite.}

\item{The \fuse\ composite is
significantly harder than the EUV portion of the \hst\ composite
spectrum of T02 who find
$\alpha=-1.76 \pm 0.12$ for 332 spectra of 184 AGNs with $z>0.33$.}

\item{We find significant correlations of EUV spectral index with
redshift and luminosity in the \fuse\ AGN sample and in the \fuse\
sample combined with the sample of T02 in the sense that lower redshift/luminosity
AGNs have harder spectral slopes.}

\item{From comparisons of the \fuse\ spectra with data from \hst\ and
\iue, we find no evidence for a UV spectral break at
$\sim$1200 \AA\ in the \fuse\ AGN sample.}

\item{We find a significant correlation of EUV spectral index with black
hole mass for 22 AGNs for which such measurements exist in the literature. The
trend of EUV spectral index with disk temperature, estimated from the 
AGN luminosities and black hole masses, runs in the expected direction.
The correlation between these quantities is not significant, perhaps due
to observational uncertainties.}

\item{The \fuse\
sample is dominated by low-redshift, low-luminosity AGNs. 
Therefore, items 3-7 suggest the
physical underpinnings of the Baldwin effect, the tendency
for lower luminosity AGNs to have hotter accretion disks and harder
spectra in the UV and soft X-rays.}

\end{enumerate}

\acknowledgements
J.\ E.\ S.\ acknowledges helpful discussions with L.\ Ho and P.\ Hall.
This research has made use of the NASA/IPAC Extragalactic Database (NED) which is
operated by the Jet Propulsion
Laboratory, California Institute of Technology, under contract with the National
Aeronautics and Space  Administration.

{}

\onecolumn
\begin{deluxetable}{lcccrccc}
\rotate
\tablecolumns{8}
\tablewidth{46pc}
\tablecaption{AGN in {\it FUSE} Sample
\label{table-alpha}}
\tablehead{
\colhead{Name} &\colhead{$z$} &\colhead{E(B-V)} &\colhead{$\log (\lambda L)$\tablenotemark{1} }
&\colhead{$\alpha_{\lambda}$\tablenotemark{2}}
&\colhead{F$_{\lambda}$\tablenotemark{2}} 
&\colhead{$\log {\rm M}_{\rm BH}/{\rm M}_{\sun}$\tablenotemark{3}} 
&\colhead{$\log (L_{\rm bol})\tablenotemark{4}$} }
\startdata
NGC 3783        & 0.010 & 0.119 &43.7 &$  1.30\pm 0.14$ &$ 2.04\pm 0.03$ &  $7.04^{+  0.42}_{-  0.39}$ &44.41 \\
Mrk 352         & 0.015 & 0.061 &43.1 &$  2.69\pm 0.42$ &$ 2.37\pm 0.01$ &\nodata & \nodata \\
NGC 7469        & 0.016 & 0.069 &43.9 &$  1.76\pm 0.09$ &$ 1.32\pm 0.01$ &  $6.88^{+  0.43}_{-  0.43}$ &45.28 \\
Mrk 79          & 0.022 & 0.071 &43.6 &$  1.17\pm 0.41$ &$ 0.29\pm 0.01$ &  $8.01^{+  0.17}_{-  0.24}$ &44.57 \\
Ark 564         & 0.025 & 0.060 &43.3 &$  1.57\pm 0.18$ &$ 0.160\pm 0.004$ &\nodata & \nodata \\
Mrk 335         & 0.026 & 0.035 &44.3 &$  1.68\pm 0.06$ &$ 1.20\pm 0.01$ &  $6.58^{+  0.16}_{-  0.11}$ &44.69 \\
Mrk 290         & 0.030 & 0.015 &43.5 &$  1.90\pm 0.23$ &$ 0.14\pm 0.04$ &\nodata & \nodata \\
Mrk 279         & 0.030 & 0.016 &44.5 &$  1.91\pm 0.05$ &$ 1.270\pm 0.007$ &\nodata & \nodata \\
Mrk 817         & 0.031 & 0.007 &44.4 &$  1.50\pm 0.08$ &$ 0.87\pm 0.01$ &  $7.55^{+  0.13}_{-  0.11}$ &44.99 \\
Ark 120         & 0.032 & 0.128 &44.5 &$  2.82\pm 0.14$ &$ 1.25\pm 0.02$ &  $8.27^{+  0.09}_{-  0.10}$ &44.91 \\
IRAS F11431-1810& 0.033 & 0.039 &44.4 &$  4.03\pm 0.18$ &$ 1.13\pm 0.02$ &\nodata & \nodata \\
Mrk 509         & 0.034 & 0.057 &44.4 &$  1.04\pm 0.28$ &$ 1.34\pm 0.05$ &  $7.96^{+  0.05}_{-  0.05}$ &45.03 \\
Mrk 618         & 0.035 & 0.076 &44.5 &$  2.51\pm 0.34$ &$ 1.20\pm 0.06$ &\nodata & \nodata \\
ESO 141-G55     & 0.036 & 0.111 &44.8 &$  3.27\pm 0.14$ &$ 2.78\pm 0.06$ &  $8.85^{+  0.67}_{-  0.43}$ & 45.62\tablenotemark{4}\\
Mrk 9           & 0.040 & 0.059 &44.3 &$  0.95\pm 0.18$ &$ 0.344\pm 0.009$ &\nodata & \nodata \\
1H 0707-495     & 0.041 & 0.095 &44.4 &$  3.63\pm 0.11$ &$ 0.86\pm 0.01$ &\nodata & \nodata \\
NGC 985         & 0.042 & 0.033 &44.2 &$  3.10\pm 0.06$ &$ 0.403\pm 0.003$ &\nodata & \nodata \\
KUG 1031+398    & 0.042 & 0.015 &43.3 &$  2.83\pm 0.27$ &$ 0.051\pm 0.001$ &\nodata & \nodata \\
Mrk 506         & 0.043 & 0.031 &44.0 &$  2.42\pm 0.24$ &$ 0.209\pm 0.007$ &\nodata & \nodata \\
Fairall 9       & 0.047 & 0.027 &43.9 &$  1.04\pm 0.33$ &$ 0.172\pm 0.008$ &  $7.92^{+  0.13}_{-  0.22}$ &45.23 \\
Mrk 734         & 0.050 & 0.032 &44.5 &$  1.37\pm 0.19$ &$ 0.54\pm 0.01$ &\nodata & \nodata \\
IRAS 09149-62   & 0.057 & 0.182 &45.0 &$  4.34\pm 0.32$ &$ 1.31\pm 0.05$ &\nodata & \nodata \\ 
ESO 265-G23     & 0.056 & 0.096 &44.4 &$  2.14\pm 0.43$ &$ 0.38\pm 0.01$ &\nodata & \nodata \\ 
3C 382          & 0.058 & 0.070 &44.4 &$  2.48\pm 0.29$ &$ 0.36\pm 0.01$ &\nodata & \nodata \\
PG 1011-040     & 0.058 & 0.037 &44.5 &$  0.67\pm 0.07$ &$ 0.401\pm 0.003$ &\nodata & \nodata \\ 
Mrk 1298        & 0.060 & 0.055 &44.5 &$  1.14\pm 0.46$ &$ 0.39\pm 0.02$ &\nodata & \nodata \\
I Zw 1          & 0.061 & 0.065 &44.2 &$ -0.51\pm 0.27$ &$ 0.157\pm 0.006$ &\nodata & \nodata \\
Ton S180        & 0.062 & 0.014 &44.9 &$  1.93\pm 0.14$ &$ 0.96\pm 0.01$ &\nodata & \nodata \\
II Zw 136       & 0.063 & 0.044 &44.6 &$  1.57\pm 0.31$ &$ 0.464\pm 0.006$ &\nodata & \nodata \\
PG 1229+204     & 0.063 & 0.027 &44.4 &$ -1.12\pm 0.58$ &$ 0.278\pm 0.007$ &  $7.93^{+  0.21}_{-  0.20}$ &45.01 \\
Ton 951         & 0.064 & 0.037 &44.8 &$  0.36\pm 0.17$ &$ 0.720\pm 0.005$ &\nodata & \nodata \\
MR 2251-178     & 0.066 & 0.039 &44.5 &$  0.69\pm 0.08$ &$ 0.299\pm 0.003$ &\nodata & \nodata \\
Ton 1187        & 0.070 & 0.011 &44.4 &$  0.84\pm 0.59$ &$ 0.223\pm 0.006$ &\nodata & \nodata \\
Mrk 205         & 0.071 & 0.042 &44.2 &$  1.76\pm 0.60$ &$ 0.107\pm 0.003$ &\nodata & \nodata \\
Mrk 478         & 0.079 & 0.014 &44.8 &$  1.29\pm 0.32$ &$ 0.443\pm 0.005$ &\nodata & \nodata \\
VII Zw 118      & 0.080 & 0.038 &44.8 &$  1.46\pm 0.30$ &$ 0.427\pm 0.004$ &\nodata & \nodata \\
PG 1211+143     & 0.081 & 0.035 &45.2 &$  2.83\pm 0.27$ &$ 0.868\pm 0.008$ &  $7.37^{+  0.10}_{-  0.13}$ &45.81 \\
Mrk 1383        & 0.086 & 0.032 &45.2 &$  0.72\pm 0.08$ &$ 1.06\pm 0.004$ &\nodata & \nodata \\
PG 0804+761     & 0.100 & 0.035 &45.7 &$ -0.53\pm 0.05$ &$ 1.010\pm 0.003$ &  $8.21^{+  0.04}_{-  0.04}$ &45.93 \\
PG 1415+451     & 0.114 & 0.009 &44.7 &$  1.88\pm 0.24$ &$ 0.155\pm 0.003$ &\nodata & \nodata \\
Ton S210        & 0.116 & 0.017 &45.4 &$ -0.57\pm 0.07$ &$ 0.790\pm 0.007$ &\nodata  &\nodata \\ 
RX J1230.8+0115 & 0.117 & 0.019 &45.3 &$ -0.24\pm 0.20$ &$ 0.63\pm 0.01$ &\nodata & \nodata \\
Mrk 106         & 0.123 & 0.028 &45.0 &$ -0.16\pm 0.22$ &$ 0.280\pm 0.005$ &\nodata & \nodata \\
Mrk 876         & 0.129 & 0.027 &45.7 &$ -0.92\pm 0.05$ &$ 1.070\pm 0.005$ &\nodata & \nodata \\ 
PG 1626+554     & 0.133 & 0.006 &45.0 &$ -0.04\pm 0.23$ &$ 0.218\pm 0.004$ &\nodata & \nodata \\ 
Q0045+3926      & 0.134 & 0.052 &45.0 &$  1.10\pm 0.31$ &$ 0.210\pm 0.005$ &\nodata & \nodata \\
PKS 0558-504    & 0.137 & 0.044 &45.5 &$  1.14\pm 0.06$ &$ 0.695\pm 0.003$ &\nodata & \nodata \\ 
PG 0026+129     & 0.142 & 0.071 &45.3 &$  1.24\pm 0.16$ &$ 0.445\pm 0.006$ &  $7.42^{+  0.08}_{-  0.09}$ &45.39 \\ 
PG 1114+445     & 0.144 & 0.016 &44.3 &$  3.44\pm 0.67$ &$ 0.033\pm 0.002$ &\nodata & \nodata \\ 
PG 1352+183     & 0.152 & 0.019 &44.7 &$  1.89\pm 0.40$ &$ 0.080\pm 0.002$ &\nodata & \nodata \\
MS 07007+6338   & 0.153 & 0.051 &45.3 &$  0.68\pm 0.16$ &$ 0.429\pm 0.004$  &\nodata & \nodata \\
PG 1115+407     & 0.154 & 0.016 &44.8 &$  0.89\pm 0.23$ &$ 0.099\pm 0.002$ &\nodata & \nodata \\
PG 0052+251     & 0.155 & 0.047 &45.2 &$  0.90\pm 0.03$ &$ 0.315\pm 0.005$ &$8.48^{+  0.13}_{-  0.11}$ &45.93 \\
PG 1307+085     & 0.155 & 0.034 &45.5 &$  0.07\pm 0.08$ &$ 0.557\pm 0.004$ &$8.52^{+  0.16}_{-  0.29}$ &45.83 \\
3C 273          & 0.158 & 0.021 &46.4 &$  0.40\pm 0.03$ &$ 3.630\pm 0.007$ &$8.54^{+  0.41}_{-  0.41}$ &47.24\tablenotemark{4}\\
PG 1402+261     & 0.164 & 0.016 &45.5 &$  0.82\pm 0.13$ &$ 0.431\pm 0.004$ &  $7.76^{+  0.91}_{-  0.30}\, $\tablenotemark{3} & 45.13 \\
PG 1048+342     & 0.167 & 0.023 &45.0 &$  1.18\pm 0.26$ &$ 0.138\pm 0.003$ &\nodata & \nodata \\ 
PG 1322+659     & 0.168 & 0.019 &45.1 &$  1.42\pm 0.29$ &$ 0.177\pm 0.004$ &\nodata & \nodata \\ 
PG 2349-014     & 0.174 & 0.027 &45.4 &$ -1.08\pm 0.14$ &$ 0.258\pm 0.003$ &  $9.26^{+  0.91}_{-  0.30}\, $\tablenotemark{3} & 45.94 \\
PG 1116+215     & 0.176 & 0.023 &45.9 &$ -0.40\pm 0.03$ &$ 0.891\pm 0.003$ &  $8.70^{+  0.91}_{-  0.30}\, $\tablenotemark{3} & 46.02 \\
FBQS J2155-0922 & 0.190 & 0.046 &46.0 &$  0.02\pm 0.12$ &$ 1.03\pm0.01$    &\nodata & \nodata \\
4C +34.47       & 0.206 & 0.037 &45.1 &$  0.12\pm 0.36$ &$ 0.099\pm 0.003$ &\nodata & \nodata \\
PG 0947+396     & 0.206 & 0.019 &45.2 &$  1.15\pm 0.25$ &$ 0.155\pm 0.003$ &\nodata & \nodata \\
HE 1050-2711    & 0.208 & 0.070 &45.2 &$  0.55\pm 0.30$ &$ 0.146\pm 0.004$ &\nodata & \nodata \\
HE 1115-1735    & 0.217 & 0.037 &45.6 &$ -0.16\pm 0.12$ &$ 0.278\pm 0.003$  &\nodata & \nodata \\
PG 0953+414     & 0.234 & 0.013 &46.0 &$  0.75\pm 0.03$ &$ 0.605\pm 0.001$ & $8.21^{+  0.07}_{-  0.08}$ & 46.16 \\
HE 1015-1618    & 0.247 & 0.078 &45.4 &$ -1.33\pm 0.34$ &$ 0.135\pm 0.004$ &\nodata & \nodata \\
PG 1444+407     & 0.267 & 0.014 &45.6 &$  0.91\pm 0.13$ &$ 0.202\pm 0.002$ & $8.54^{+  0.91}_{-  0.42}\, $\tablenotemark{3} & 45.93 \\
PKS 1302-102    & 0.278 & 0.043 &45.8 &$  0.09\pm 0.05$ &$ 0.280\pm 0.001$ &  $8.78^{+  0.94}_{-  0.30}\, $\tablenotemark{3} & 45.86 \\
HE 0450-2958    & 0.286 & 0.015 &45.5 &$  0.79\pm 0.13$ &$ 0.157\pm 0.002$ &\nodata & \nodata \\
PG 1100+772     & 0.311 & 0.034 &45.7 &$  0.22\pm 0.15$ &$ 0.238\pm 0.005$ &\nodata & \nodata \\
Ton 28          & 0.330 & 0.022 &45.9 &$  1.97\pm 0.13$ &$ 0.307\pm 0.005$  &\nodata & \nodata \\
FBQS J083535.8+24 & 0.331 & 0.031 &45.7 &$  2.42\pm 0.25$ &$0.188\pm 0.005$  &\nodata & \nodata \\
PG 1216+069     & 0.331 & 0.022 &45.7 &$  1.20\pm 0.18$ &$ 0.184\pm 0.004$ &\nodata & \nodata \\
PG 1512+370     & 0.371 & 0.022 &45.6 &$  2.77\pm 0.53$ &$ 0.137\pm 0.009$ &\nodata & \nodata \\ 
PG 1543+489     & 0.400 & 0.018 &45.7 &$  0.83\pm 0.54$ &$ 0.099\pm 0.007$ &\nodata & \nodata \\ 
FBQS J152840.6+28 & 0.450 & 0.024 &46.0 &$  0.61\pm 0.19$ &$ 0.177\pm 0.004$  &\nodata & \nodata \\
PG 1259+593     & 0.478 & 0.008 &46.1 &$    0.42\pm 0.02$ &$ 0.1810\pm 0.0004$ &\nodata & \nodata \\
HE 0226-4110    & 0.495 & 0.016 &46.4 &$    0.70\pm 0.04$ &$ 0.345\pm 0.001$   &\nodata & \nodata \\
HS 1102+3441    & 0.510 & 0.024 &46.0 &$    2.83\pm 0.24$ &$ 0.154\pm0.003$    &\nodata & \nodata \\
RX J2241.8-44   & 0.545 & 0.011 &46.0 &$    3.47\pm 0.16$ &$ 0.122\pm0.002$    &\nodata & \nodata \\
PKS 0405-12     & 0.573 & 0.058 &46.9 &$ 3.35\pm0.08$   & $ 1.13 \pm0.01$    &\nodata & \nodata \\ 
HE 1326-0516    & 0.578 & 0.030 &46.4 &$-0.33\pm0.13$   & $ 0.141\pm0.002$   &\nodata & \nodata \\
HE 0238-1904    & 0.631 & 0.032 &46.5 &$1.61\pm0.11$    & $ 0.287\pm0.003$   &\nodata & \nodata \\ 
3C 57           & 0.669 & 0.022 &46.4 &$1.52\pm0.14$    & $ 0.169\pm0.003$   &\nodata & \nodata \\ 
\enddata
\tablenotetext{1}{Luminosity at 1100 \AA, $\log (\lambda L_{1100})$ 
for $\Omega_{0}=1$ and $H_{0}=60$ km~s$^{-1}$ Mpc$^{-1}$, see Equ.~\ref{equ:lum}}
\tablenotetext{2}{Spectral index, $F_\lambda \propto \lambda^{-\alpha_{\lambda}}$, and flux 
at 1000 \AA\ in units of $10^{-13}$ ergs s$^{-1}$ cm$^{-2}$ \AA$^{-1}$ from power law continuum fit}
\tablenotetext{3}{Measurements of ${\rm M}_{\rm BH}$ from Kaspi et al.\ (2000) unless marked with $``$3",
which are from M$^{\rm c}$Lure \& Dunlop (2001)}
\tablenotetext{4}{Bolometric Luminosity calculated by Woo \& Urry 2002 unless marked with
 $``$4", which are from Padovani \& Rafanelli (1988)}
\end{deluxetable}

\begin{deluxetable}{lccr}
\tablecolumns{4}
\tablewidth{24pc}
\tablecaption{Fits to Full Sample and Subsample Composites \label{table-subsamples}}
\tablehead{
\colhead{Sample} &\colhead{No.\ Spectra}
&\colhead{No.\ AGNs} &\colhead{$\alpha$}}
\startdata
Full sample                      &128 &85 &-0.56\\
$z<0.10$                         &65 &39 &-0.74 \\
$z>0.10$                         &63 &46 &-0.74 \\ 
$\log(\lambda L_{1100}) < 45.0$  &64 &40 & 0.19 \\
$\log(\lambda L_{1100}) > 45.0$  &64 &45 &-0.84 \\
$E(B-V)<0.032$                   &65 &44 &-0.50 \\
$E(B-V)>0.032$                   &63 &41 &-0.21 \\
\enddata
\end{deluxetable}

\begin{deluxetable}{llcc}
\tablecolumns{4}
\tablewidth{23pc}
\tablecaption{Emission Lines
\label{table-emspec}}
\tablehead{
\colhead{Line} &\colhead{$\lambda_{0}$ (\AA)} 
&\colhead{Flux}
&\colhead{EW (\AA)} }
\startdata
Ne{\sc viii} + O{\sc iv}     &772     &6.3 $\pm$0.3 &8.4 \\ 
O{\sc iii}                   &831     &1.6 $\pm$0.3 &2.4 \\
\hi\ Ly series + S{\sc vi}   &944     &3.2 $\pm$0.5 &5.6 \\
Ly$\gamma$                   &973     &1.1 $\pm$1.1 &2.0 \\
C{\sc iii}                   &976     &1.5 $\pm$0.4 &2.9 \\
N{\sc iii}                   &991     &1.1 $\pm$0.2 &2.1 \\
Ly$\beta$                    &1026    &6.2 $\pm$1.3 &12.6 \\
O{\sc vi}                    &1033    &10.0$\pm$0.3 &20.3 \\
S{\sc iv}                    &1062    &1.7 $\pm$0.3 &3.6 \\
S{\sc iv}                    &1073    &0.8 $\pm$0.1 &1.8 \\ 
N{\sc ii} + He{\sc ii} + Ar{\sc i} &1084  &0.23$\pm$0.05 &0.4 \\
\enddata
\end{deluxetable}

\begin{deluxetable}{llrrrrrr}
\tablecolumns{8}
\tablewidth{38pc}
\tablecaption{Fits to Individual AGN \label{table-linestats}}
\tablehead{
\multicolumn{4}{c}{} &\multicolumn{2}{c}{\underline{$\alpha$ vs. $\log z$}} 
&\multicolumn{2}{c}{\underline{$\alpha$ vs. $\log \lambda L$}} \\
\colhead{Sample} &\colhead{N} 
&\colhead{$\alpha_{\rm mean}$} &\colhead{$\alpha_{\rm med.}$}
&\colhead{Slope} &\colhead{r$_{\rm s}$} 
&\colhead{Slope} &\colhead{r$_{\rm s}$}}
\startdata
{\it FUSE}       &85  &-0.74$\pm$0.13 &-0.86$\pm$0.13 &-0.83$\pm$0.02 &-0.31
                 &-0.56$\pm$0.01 &-0.36  \\
{\it FUSE + HST} &164 &-1.25$\pm$0.09 &-1.40$\pm$0.09 &-0.88$\pm$0.13 &-0.45
                 &-0.49$\pm$0.09 &-0.38 \\
\enddata
\end{deluxetable}

\clearpage

\clearpage

\clearpage

\clearpage

\clearpage

\clearpage

\clearpage

\clearpage

\clearpage

\clearpage

\clearpage

\clearpage

\clearpage

\clearpage

\clearpage

\clearpage

\end{document}